\documentclass[11pt,a4paper]{article}
\usepackage[a4paper,margin=25mm]{geometry}
\usepackage{lmodern}
\usepackage{microtype}
\usepackage{amsmath}
\usepackage[hyphens]{url}
\usepackage{lastpage}
\usepackage{fancyhdr}

\usepackage{graphicx}
\usepackage{booktabs}
\usepackage{hyphenat}
\usepackage[all]{nowidow}
\usepackage{tabularx} 
\usepackage{floatrow} 
\usepackage{refcount} 
\usepackage[noadjust]{cite} 

\usepackage{xcolor}
\usepackage{hyperref}
\hypersetup{
    urlbordercolor = black,
    pdfborderstyle={/S/U/W 1},
}

\usepackage{cleveref} 

 \usepackage{array}
\usepackage{longtable}
\usepackage[T1]{fontenc}
\ifdefined\pdfgentounicode
\input{glyphtounicode}
\fi
\newcommand\eIbrk{\discretionary{}{}{}}
\newcommand\eIcode[1]{\begingroup\ttfamily\let\_\textunderscore
  \nohyphens{#1}\endgroup}

\hypersetup{pdftitle={Where Accountability Lives: Mapping Human Responsibility to Workflow Artifacts in Agentic Software Development},pdfauthor={Sabry E. Farrag; Youssef Saber}}
\makeatletter
\newcommand\eIkeepwithcaption[1]{%
  \penalty-100
  \begingroup
    \setlength{\dimen@}{#1}%
    \vskip\dimen@\penalty9999\vskip-\dimen@\kern\z@
  \endgroup}
\newcommand\eIkeepwithtable[1]{\penalty-100}
\makeatother
\begin{document}
\pagestyle{fancy}\fancyhf{}\renewcommand{\headrulewidth}{0pt}
\fancyhead[L]{\small Farrag and Saber, Where Accountability Lives (preprint v2)}\fancyhead[R]{\small\thepage\ of \pageref{LastPage}}
\newcommand{\orcidmark}[1]{\begingroup\hypersetup{pdfborderstyle={},pdfborder={0 0 0}}\href{https://orcid.org/#1}{\raisebox{-0.2ex}{\includegraphics[height=1.7ex]{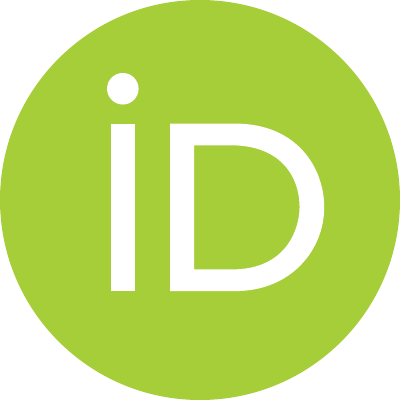}}}\endgroup}
\title{Where Accountability Lives: Mapping Human Responsibility to Workflow
Artifacts in Agentic Software Development}
\author{\begin{minipage}[t]{0.47\textwidth}\centering Sabry E. Farrag\,\orcidmark{0009-0002-0735-7443}$^{*}$\\[1pt] {\small School of Architecture, Computing and Engineering, University of East London, London, United Kingdom}\\[1pt] {\small\texttt{u3080723@uel.ac.uk}}\end{minipage}\hspace{0.02\textwidth}\begin{minipage}[t]{0.47\textwidth}\centering Youssef Saber\,\orcidmark{0009-0008-4028-1587}\\[1pt] {\small School of Electronics, Electrical Engineering and Computer Science, Queen's University Belfast, United Kingdom}\\[1pt] {\small\texttt{yemanmohammedsaber01@qub.ac.uk}}\end{minipage}\\[2pt] {\small $^{*}$Corresponding author.}}
\date{}
\maketitle
\begin{abstract}
\noindent \textbf{Context}:
Coding agents author commits, open pull requests and push code in production
repositories. Between provider and user, responsibility is settled in two
layers that do not refer to each other: platform controls gating what an agent
may do, and provider terms allocating responsibility for its output. \
\textbf{Objective}:
To establish where the two layers disagree at a workflow event, and what each
states there about authority, execution, verification, consequence and record. \
\textbf{Method}:
A qualitative document study of 121 items archived byte-exact with
provenance headers: documentation for four agentic coding tools, the platform
controls acting on their output, and eighteen policy documents from seven
providers. Deductive content analysis against an a-priori system of
five dimensions and nine workflow events. An independent human second
coder blind-recoded the verification-mechanism classification, a complete
enumeration and not a sample: Cohen's $\kappa$ is 0.81 ($n = 13$) on whether a mechanism compels,
0.75 ($n = 12$) on who performs it. Retrieval and extraction run
as deposited scripts, no language model in the path. \
\textbf{Results}:
Cursor's terms make the user responsible for evaluating the use of any
suggestion; its documentation describes a product that performs that
evaluation and records it as an approval. In the collected documentation no
artifact records that the verification duty the terms impose on the user was
discharged; GitHub's audit log records a review's submission but not its state.
Verification has two axes, whether a mechanism compels the check and whether a
person performs it. At merge the tools take three positions: two compel a
person, one barring the agent from merging its own pull request and one
requiring a person's retry in auto mode; one documents an agent approving
below a configured risk threshold; one only advises. Six of the fifteen
compelling mechanisms have a default the vendor states; nine have one this
study inferred. Readers who had not made them refuted five of the eight
absence claims; three survived, one materially qualified. \
\textbf{Conclusions}:
The terms attach duty and consequence to output as a class, the platform
records events, nothing records the duty discharged. One vendor documents
a product that forms the approval judgement, and its terms make the user
answerable for the output regardless. Peer-reviewed repository-scale
measurements of agent-authored pull requests are consistent with that record
gap.
\end{abstract}


\noindent\textbf{Dataset:} \url{https://doi.org/10.5281/zenodo.21965181}

\noindent\textbf{Version note:} Preprint, version 2, September 2026, revised throughout from version 1 of 16 August 2026 (arXiv:2608.15678v1): the analysis is restated as a deductive qualitative content analysis, a second and a third coder were added, the absence claims were re-tested and five were refuted, and the source collection was re-deposited. This version has not been peer reviewed.

\section{Introduction}\label{sec:Introduction}

A coding agent is assigned an issue. It creates a branch, writes code, commits it under its own name, and opens a pull request. A developer reads the diff, approves it, and merges. The change later fails in production. Who was responsible for catching it, and what record exists of whether anyone tried?

Two layers answer that question, and they do not refer to each other. Platform controls set what an agent may do and which human actions must precede it: branch protection rules, required approvals, workflow execution gates, deployment reviewers. Provider terms allocate responsibility for what the agent produces, as between provider and user. The first layer is enforced by software, the second by contract, and the two are written at different levels. Documentation describes events in a repository; terms describe output as a class. A later reader may find an approval, a merge or an audit-log entry and still be unable to say whether the verification the terms require took place.

Treude~\cite{Treude2026} read the terms of service of nine AI coding tools and found providers granting output rights to users while allocating responsibility for correctness to them, regardless of how much autonomous action produced the output. He proposed a direction rather than taking it: connect responsibility to specific artifacts and events in the workflow, so that accountability is represented during normal practice rather than reconstructed after a failure. This paper takes that step. It reads the contractual layer against the platform layer operating on the same workflow, and anchors both to the events that leave observable traces in a repository.

\subsection{Research questions}\label{sec:ms1.1}

Three questions organise the study, each answerable from published material, and they are asked in the order the argument takes them: whether the two layers agree, what each states that makes the comparison possible, and what the events leave behind.

\textbf{RQ1.} Where the two layers both speak to an event, do they agree, and where they disagree, what is the shape of the disagreement?

\textbf{RQ2.} At each workflow event where an agent acts or a human gates it, what do vendor documentation and provider terms state about authority, execution, verification, consequence, and record? Two layers cannot be compared at an event until what each states there is recorded, and the Accountability Event Map is that record.

\textbf{RQ3.} Which of these events produces a durable artifact from which a later question about responsibility could be answered?

\subsection{The thesis}\label{sec:ms1.2}

An approval on a pull request records that a party holding authority performed an action the platform stores as approval. It does not record what that party understood. The gap between the act and the artifact was documented a decade ago. Bacchelli and Bird~\cite{Bacchelli2013} card-sorted 570 review comments at Microsoft and ranked defects fourth of nine categories, at 14 per cent, behind code improvements at 29 per cent. Czerwonka et al.~\cite{Czerwonka2015}, under the title \emph{Code Reviews Do Not Find Bugs}, put possible defects at about 15 per cent of reviewer comments against at least 50 per cent for maintainability feedback. Sadowski et al.~\cite{Sadowski2018} describe a process in which over 80 per cent of changes involve at most one iteration of resolving comments and 70 per cent are committed less than 24 hours after being mailed out for an initial review. All three are single-company, so together they fix a direction rather than a measurement of the field. The closest measurement is McIntosh et al.~\cite{McIntosh2014,McIntosh2016}: across Qt, VTK and ITK, components whose changes were reviewed with low participation, including changes approved without discussion, had more post-release defects. Reviewers throughout were forming judgements whose content the artifact recorded none of.

\textbf{Agentic tooling did not create an accountability gap.} The approval artifact recorded less than a reader might assume long before any agent existed, and the terms in Section~\ref{sec:ms4.4} attach consequence to output as a class. What is new is narrower. One vendor's documentation now describes a design in which the party that forms the approval judgement and emits that artifact is not a person, under a threshold a human configured once. Section~\ref{sec:ms4.2} gives the documentation, Section~\ref{sec:ms5.1} what separates that design from the rule-following bots that have occupied the same event for a decade, and Section~\ref{sec:Threats} a second vendor documenting the same arrangement since this study's retrieval date. The claim is about those vendors on those dates, not about the class of tools.

\subsection{What this study did, and what it contributes}\label{sec:ms1.3}

This study collected the official product documentation for four widely used agentic coding tools, the platform controls that apply to their output, and eighteen governing policy documents that allocate responsibility for it. It then mapped responsibility onto nine workflow events, from task assignment to deployment. At each event it recorded the \textbf{five accountability dimensions}: authority, who may act; execution, who acted and under whose identity; verification, who is obliged to check and whether a mechanism compels it; consequence, which party the terms allocate it to; and record, which artifact shows the event occurred. The result is the \textbf{Accountability Event Map}. Table 1 prints one tool's slice of it, and the map for all four is deposited as Appendix A (\eIcode{appendix\_A\_evidence\_map.md}, Section~\ref{sec:ms3.5}) and cited here by cell, A.1 to A.9.

The paper makes four contributions. The first is \textbf{two documented cases} in which one provider's product documentation and governing terms do not meet: in one, the terms make the user responsible for evaluating the use of any suggestion while the documentation describes a product that performs that evaluation and records it as an approval; in the other, the documentation describes an autonomous agent and the terms a return value. The second is \textbf{a mechanism-by-actor grid for verification}. Governance frameworks read verification on one scale, their in-, over- and out-of-the-loop tiers, and a case in this corpus could not be placed on it. Whether a mechanism compels verification and whether the performer is a person are separate questions, and the grid crosses them. The third is \textbf{the map itself}, deposited whole, so that a reader who rejects a cell can re-code it against the archived body. The fourth is \textbf{an archived source collection} underlying every claim about a collected document, with the coding frame as a standalone codebook, the completed second coding and the harness that scores it, and the deposited scripts.

Two questions this paper does not answer belong to a study of repository and incident data: what practitioners do at these events, and what the correspondence gap costs anyone. Section~\ref{sec:ms8.1} states the bounds and the Conclusion what such a study would need.

\subsection{How to read this paper}\label{sec:ms1.4}

Every claim about a collected document carries the archive item identifier of the page it rests on. The wording is deposited rather than printed: the evidence map of Appendix A holds the quoted wording behind every cell that rests on one (Section~\ref{sec:ms3.5}).

Four displays carry the result, three of them read off the map's cells. Figure~\ref{fig:vocabgrid} records which of the twenty-one terms of verification duty each instrument uses, and is the one display that rests on the term search over the policy instruments rather than on the map. Figure~\ref{fig:eventmap} prints the map itself. Figure~\ref{fig:grid} puts fifteen verification mechanisms on the grid that separates whether a mechanism compels a check from whether a person performs it, and shows one provider in the cell where the mechanism compels and the performer is not a person. Figure~\ref{fig:pipeline} sets the nine events against what each layer states at them. Table 3 gives the eighteen mechanisms, fifteen of which compel a check.

The rest runs in the venue's order, with two departures: Section~\ref{sec:Cases} states the two documented cases before the map, and Section~\ref{sec:Results} puts RQ1 to the map dimension by dimension, its Record subsection answering RQ3.

\section{Background}\label{sec:Background}

The problem this paper observes at the level of the commit has a long prior life at the level of the person. Bovens~\cite{Bovens2007} gives the relational definition the governance literature uses: an actor, a forum, an obligation to explain and justify, the forum's right to question and judge, and consequences. The five dimensions of Section~\ref{sec:ms3.2} decompose the actor's side as a repository can observe it; the forum is not in the corpus, which is why Section~\ref{sec:ms4.3} reports durability horizons without a threshold. Matthias~\cite{Matthias2004} states the general form of what follows: a responsibility gap arises where a system's behaviour is not fully predictable by its operator, so the conditions for ascribing responsibility to a human are unmet while the machine cannot bear it either. Elish~\cite{Elish2019}, the closest antecedent, describes a moral crumple zone, in which responsibility for a failure is absorbed by the nearest human operator whose actual control was limited. Section~\ref{sec:ms5.2} finds that structure in the documentation.

\section{Related work}\label{sec:RelatedWork}

Accountability in agent-mediated software development has been examined at three levels, each with its own unit of analysis. None of them is the workflow event.

\subsection{Contracts and governance}\label{sec:ms2.1}

Treude~\cite{Treude2026} analysed fourteen policy documents across nine AI coding assistants and agent-enabled tools, coding clauses along ownership, responsibility and liability, data governance, and delegation framing. The architecture was consistent: providers assign output rights to users while placing responsibility for correctness, legality and downstream consequences on them, an allocation coherent for assistive use under close supervision and strained as agents plan and execute larger changes. That study establishes what the contracts say, without comparing them against the platform mechanisms operating on the same work. This paper's collection overlaps that corpus and follows its delegation chains out to the documents those instruments point to. The difference is what reading one layer cannot reach: the instrument allocates the responsibility without naming the mechanism that would discharge it, the documentation describes a product that performs that evaluation and records the result as an approval without relating it to the allocation, and the conflict between them is stated in neither and appears only at the event where both apply.

Governance frameworks allocate accountability across organisational roles instead. The Singapore framework, version 1.0, makes human accountability one of four governance pillars and adapts human-in-the-loop models to counter automation bias~\cite{IMDA2026}. Kang~\cite{Kang2026GAIE} routes a code generation task to one of three oversight tiers and specifies for each the evidence artifacts a compliance audit would require. Hydari and Muzaffar~\cite{Hydari2026} hold that agentic AI redraws organisational boundaries around accountability rather than dissolving them, and that verification cost and responsibility transferability determine whether execution and accountability boundaries can move together. The unit in each is the role or the firm, not the artifact, and each prescribes what a governed workflow should produce. The question here is its complement: what four vendors and seven providers currently document.

Regulation makes the same move at a higher level of generality. Article 14 of Regulation (EU) 2024/1689, the Artificial Intelligence Act, requires high-risk AI systems to be designed so that natural persons can effectively oversee them in use, and Article 26(2) requires the deploying organisation to assign that oversight to natural persons with the competence, training and authority to exercise it~\cite{EUAIAct2024}, the legal analogue of this paper's Verification row. Two words collide between the vocabularies, and this paper keeps its own sense of both: its \emph{provider} is the vendor that publishes the tool and the instrument, and its \emph{user} the customer bound by it, the Regulation's \emph{deployer}. The reach of those duties here is conditional, nothing in the Regulation's classification placing agentic coding tools among high-risk systems.

\subsection{Review, provenance, and auditability}\label{sec:ms2.2}

Dorner et al.~\cite{Dorner2025}, Kamal{\i} et al.~\cite{Kamali2026}, Velasco et al.~\cite{Velasco2026}, Raees and Papangelis~\cite{Raees2026} and Ferino et al.~\cite{Ferino2026} each treat accountability as context for a different contribution: how developers expect review to evolve, reviewers as supervisory operators of agents, provenance of generated code as a traceability problem, the measurement constructs for appropriate reliance on AI, and a reliance-control framework keyed to developer control. None provides the mapping: which workflow event carries which obligation, under which layer, and with what artifact to show for it. A separate line secures the provenance of software artifacts rather than the allocation of judgement over them. In-toto (Torres-Arias et al.~\cite{TorresArias2019}), SLSA~\cite{SLSA2026} and Sigstore (Newman et al.~\cite{Newman2022}) establish that a step ran and that a named identity ran it. None establishes that a person reviewed the change.

\subsection{Event-level accountability and bots}\label{sec:ms2.3}

Three 2026 papers work at or near this paper's level, and all three reason about what a system ought to record rather than about what vendors currently document. Romanchuk and Bondar~\cite{Romanchuk2026} name the condition this paper documents its mechanisms for: a decision occurred and no entity holds both authority over it and capacity for it, a responsibility vacuum, whose practice they call ritual review, in which the reviewer's approve-or-reject action is decoupled from their understanding. Their named instances include merge and deploy, both this paper's events, and their analysis is organisational and carries no empirical data. Nian et al.~\cite{Nian2026} decompose auditability into five dimensions with metrics, among them a responsibility attribution asking whether a delegation chain can be reconstructed at runtime; those are properties of a runtime log, and their evidence reaches no development workflow and no terms of service. Rhodes and Kang~\cite{Rhodes2026} specify a proof of execution binding a contract, a tamper-evident causal event stream and a replay context into an object a validator can check. It is not a record that a human verified anything, and the paper says so: a satisfied proof attests that execution stayed within the contract and is silent on whether that contract was well chosen.

Automation has been producing artifacts at these events for a decade, and software engineering has measured it. Wessel et al.~\cite{Wessel2018} found 93 of 351 popular GitHub projects, 26 per cent, using bots, with no consistent significant difference in project metrics before and after adoption. A bot acts through the pull request model, so its artifact has the same shape as a person's: the Execution and Record problem of Section~\ref{sec:Results} in its earlier form. Bots that open the diff are older than agents. Balachandran~\cite{Balachandran2013} placed a static-analysis review bot at the review event, and Alfadel et al.~\cite{Alfadel2021} found 65 per cent of Dependabot security pull requests accepted across 2,904 JavaScript projects. Erlenhov et al.~\cite{Erlenhov2020} characterise development bots, Wessel et al.~\cite{Wessel2022} measure that adopting a code review bot increases monthly merged pull requests, He et al.~\cite{He2023} report developers receptive to dependency bots, and Fatima et al.~\cite{Fatima2026} analyse reviewer-bot comments on agent-authored pull requests. Section~\ref{sec:ms5.1} draws the boundary this paper's documents support: not along the judgement, which several of these bots form, but along the artifact.

\subsection{Position of this paper}\label{sec:Position}

Each of those lines of work stops at the edge of the other. This paper reads the two published layers against each other at the events that produce artifacts in a repository. Its unit of analysis is the workflow event, and its question is what the published material leaves recoverable after that event: who was permitted to act, who was recorded as acting, what verification was compelled, which party bears the consequence, and what artifact survives.

\section{Method}\label{sec:Method}

Lexical rules built the corpus and shortlisted candidate passages. No language model sits anywhere in that path, and the rules decided nothing: every classification was made, or checked and adopted, by a person reading an archived document against a category system fixed in advance, and the classification behind Table 3 was re-coded independently and blind by a second person (Section~\ref{sec:Threats}). The unit of evidence is what a provider published. Every finding takes the form that the documentation states something, and every absence is an absence of documentation within a named page set.

The collection is a qualitative document study in the sense of Bowen~\cite{Bowen2009}, and the analysis a deductive, concept-driven qualitative content analysis after Schreier~\cite{Schreier2012} and Mayring~\cite{Mayring2014}, with an a-priori category system, defined coding units, written decision rules with boundary cases, and a defined marker for absence. The three patterns and the two documented cases are a thematic synthesis across the map, following Cruzes and Dyb{\aa}~\cite{CruzesDyba2011}. Reporting follows the ACM SIGSOFT Empirical Standards for Software Engineering~\cite{ralph2021empirical}, read as served on 7 September 2026~\cite{EmpiricalStandardsLiving2026}, whose General Standard directs that published guidance be applied where no standard covers the design, none of its nineteen covering document analysis.

\subsection{Design and corpus}\label{sec:ms3.1}

Two classes of document were collected. First, the official product documentation for four agentic coding tools: GitHub Copilot cloud agent, Devin, Cursor and Claude Code. Second, the governing policy documents that allocate responsibility for their output, comprising eighteen terms of service, customer agreements, and product-specific terms across seven providers. Added to these are the platform controls that operate on agent output whichever tool produced it: branch protection rules and rulesets, code owners, deployment environments, commit signature verification, and the organisation audit log, fourteen archived items over twelve distinct pages, \eIcode{GH-CTRL-01} to \eIcode{-13} and \eIcode{GIT-TRAILER-01}.

Only publisher-served documentation was collected. Blog posts, changelogs, community discussions, and third-party summaries were excluded. GitHub's artifact attestation and build provenance pages are not in the collection and no finding reaches them: the platform-control list was drawn from the controls that gate a pull request.

The archive is the union of two manifests sharing no item identifier: seventy-one items selected purposively before the coverage objection of Section~\ref{sec:ms3.3}, and fifty by the equal-effort rule adopted in answer to it, one selection rule applied to all four tools alike, so that 121 = 71 + 50, both ids kept where the rule re-selected a page already held. Four units of analysis are distinguished, from the archived document down to the single coded cell, and the deposited appendix tabulates all four with their counts. Appendix A, Section~\ref{sec:Patterns} and the answers to the research questions are claims about the thirty-six tool-event pairs, which Section~\ref{sec:Threats}'s agreement statistic does not cover.

\subsection{The coding frame}\label{sec:ms3.2}

The category system is concept-driven and has three levels: five accountability dimensions, two axes that decompose one of them, and a defined marker for absence. Appendix B gives the dimensions with an anchor each, and the frame is deposited in full as a standalone codebook. The term \emph{accountability} is stipulative here, fixed before the corpus was read: the recoverable assignment, after the event, of those five, and nothing besides.

\textbf{The five dimensions} are the rows of the map, and Table B1 gives each with a corpus anchor. \textbf{Authority}, who is permitted to cause this event. \textbf{Execution}, who performed it and under which recorded identity. \textbf{Verification}, who is obliged to check, whether a mechanism compels the check before work proceeds, and which party performs it. \textbf{Consequence}, which party the governing instrument allocates responsibility for the output to, as between provider and user, and nothing more: not third parties, not statutory limits, and not the duties background law supplies whether or not an instrument recites them (Section~\ref{sec:ms6.2}). \textbf{Record}, which artifact shows that the event occurred, and whether it is durable. \emph{Durable}, the property RQ3 asks about, reports a horizon rather than a score, in three values: \textbf{repository-durable}, with no vendor-stated expiry; \textbf{log-durable with a stated horizon}; and \textbf{session-scoped}, bound to a session, a run or a credential lifetime. Each horizon is reported as the vendor states it.

\textbf{The two axes.} Verification is the only dimension that decomposes, and its last two clauses are the axes on which every mechanism in Table 3 was coded. \eIcode{compels} asks whether a mechanism withholds progress until the check occurs; the decision rule is whether the product refuses to proceed without it. \eIcode{actor} asks which party performs it. Both are binary nominal with closed label sets, \eIcode{Yes}/\eIcode{No} and \eIcode{Human}/\eIcode{Agent}, and abstention is forbidden; an absence claim has no acting party and is coded on \eIcode{compels} alone. Table B2 fixes in advance the boundary case for a gate that can be turned off or bypassed, and the archived document as the source of truth. In the map, \eIcode{compels} admits one further value for the absence case, \eIcode{Not documented}.

\textbf{Three rules run across the frame.} \textbf{C-1}, granularity: Consequence is coded once per provider, from the governing terms, and not per event, because no collected instrument allocates responsibility at event granularity. \textbf{C-2}, applicability: at each event the dimensions coded are those the event can instantiate, a dimension not shown being judged inapplicable rather than left uncoded. \textbf{B-1}, the mode-scoped gate: a mechanism that withholds progress within an operating mode is coded \eIcode{compels = Yes} even where entering the mode is optional, Cursor's plan mode and Claude Code's permission prompt being the coded instances. The three were formalised in writing after the coding, from the coded data, and the brief of Table B2, under which Section~\ref{sec:Threats}'s agreement statistic was measured, carries the two axes but not them. The rule holds four mode-scoped rows of Table 3 at \eIcode{compels = Yes}, R7, R8, R16 and R17, so a reader who rejects it as formalised too late reads that table as eleven mechanisms compelling rather than fifteen; no other row and no other count in this paper turns on it. The full operational definitions, anchors, boundary rules, the event-by-dimension applicability matrix and the glossary for readers outside software engineering are deposited with the evidence map of Appendix A.

\textbf{NOT \nohyphens{DOCUMENTED}.} A cell takes the marker only after a person has read the archived bodies of the page set named in it. It means that no page among those searched carries a statement of the kind the row asks for, not that the behaviour does not exist. The \eIcode{Default} column of Table 3 and Table 2's summary lie outside the frame: neither was a dimension of the coding sheet and neither was second-coded.

The five dimensions were fixed before the corpus was first read and did not change. What changed sits one level below them, in six episodes deposited as section 10 of \eIcode{coding/\eIbrk{}CODEBOOK.md}, the weightiest being the replacement of the three-state verification vocabulary by the crossed axes, under a case from \eIcode{CURSOR-DOC-08} the three states could not express. A seventh changed a table rather than an instrument, and Section~\ref{sec:ms6.2} records it.

\subsection{Collection, retrieval, and absence}\label{sec:ms3.3}

Every one of the 121 manifest items was retrieved by a direct request to its origin server under the archiver's own user agent, with no fallback path, and every one returned HTTP 200. Twenty-three resolved through redirects. Each archived file carries a header giving the requested and served URLs, the HTTP status, the page title, any date string in the served markup, and the retrieval date. The 121 items are not 121 distinct pages: fourteen pairs are byte-identical, which leaves 107 distinct documents. Twelve arise from the equalisation below; the other two from URL handling, \eIcode{GH-CTRL-02} requesting the same page as \eIcode{GH-CTRL-01} with a fragment identifier never sent to the server, and \eIcode{GH-CTRL-13} redirecting onto the page already held as \eIcode{GH-CTRL-12}. Two retrieval defects were caught and corrected: three wrong GitHub URLs that returned HTTP 404, and a tag-stripping step that truncated a commit trailer. All three URLs were located, re-retrieved and their claims confirmed verbatim against the re-retrieved bodies, and the items affected by the tag-stripping step were re-fetched unprocessed. The page-by-page account of both corrections is deposited with the evidence map of Appendix A, and the consistency checker now fails on any manifest item that did not retrieve, cited or not.

\textbf{The lexical rules shortlisted wording and decided nothing.} One rule was applied to the product documentation: extract, verbatim, every sentence stating who may perform an action, whose approval is required, or what a human must do before something proceeds. Three were applied to the policy documents, on verification duties, on the allocation of responsibility or liability, and on autonomous or delegated operation. The three find what an instrument says, not what a party owes, so every result from them is a result about drafting. The rules emit every hit verbatim with its item identifier, and they returned 226 passages over the sixty-two purposive items archived when the extraction was made; the category and provenance breakdown is deposited with the extraction output. The nine purposive items added afterwards (\eIcode{GH-AGENT-04}, \eIcode{GH-AGENT-07}, \eIcode{GH-AGENT-08}, \eIcode{POL-GH-04}, \eIcode{POL-GH-06}, \eIcode{POL-GH-07}, \eIcode{POL-GH-08}, \eIcode{POL-SG-03}, \eIcode{POL-SG-04}) and the fifty equal-effort items were not put through them, and material with no hit still entered the map, by reading.

Five decisions are analytic, and no lexical rule made any of them: whether a passage describes a gate or a suggestion, which party performs an action, which cell of the map a mechanism belongs in, whether a cell is documented at all, and what a mechanism's default state is. Each was made against the frame of Section~\ref{sec:ms3.2}. The twenty-one verification-duty and nineteen autonomy terms are deposited in \eIcode{density.py} and \eIcode{absence\_check.py}, the first also transcribed with the evidence map of Appendix A, and the sixteen accountability path terms in \eIcode{equalize\_corpus.py}.

\textbf{Absence is a coded category} with the adjudication rule of Section~\ref{sec:ms3.2}. The deposited register is 440 term-by-document searches, twenty-one verification-duty and nineteen autonomy terms against each of eleven archived policy documents, of which 379 returned zero and are deposited as zeros with their surrounding text.

\textbf{The selection rule was equalised after a coverage objection.} The first collection searched five, six, nine and sixteen pages across the four tools, a spread of 3.2 to 1 in how hard each vendor had been searched, so that a difference between two tools could not be told apart from a difference in the search that found it. One rule, the equal-effort rule, now applies to all four, deposited as \eIcode{equalize\_corpus.py}: candidate pages are drawn by link extraction from each vendor's own agent-documentation index, filtered to English pages whose path matches any of sixteen accountability terms, ranked by hit count, and capped at sixteen per vendor. The final page sets are GitHub 24, Devin 14, Cursor 32, and Claude Code 19 as item counts, and 22, 12, 26 and 17 on distinct documents, the rule having re-selected pages already held under a purposive id. The residual spread of 2.17 to 1 is what Devin and Claude Code publish under the rule, their candidate sets having been taken whole where GitHub's 147 and Cursor's 23, live-network observations not reconstructible from the archive, were cut to the cap. The two spreads are not the same quantity and no threshold is asserted for either: the first measured unequal effort, which one rule removes, and the residual measures what the vendors publish under equal effort, which is a property of the corpus and is reported rather than corrected. Exhaustion here is exhaustion of the rule's candidate set, not of what a vendor publishes.

One incident in building that register is worth recording, because it is the failure mode the register is about. The first run of the survival test reported an absence refuted. The refutation was spurious: the pattern written to detect ``approve PR'' matched, case-insensitively and without a word boundary, the phrase ``approve protected-path''. Reported, it would have made this paper's headline methodological statistic wrong in the direction of appearing more self-critical than the evidence warranted, which is an error a reader is unlikely to challenge and therefore unlikely to catch. The pattern is now anchored and the test script deposited.

\textbf{Five of the eight absence claims were refuted by readers who had not made them.} A third coder's read of each vendor's full page set refuted four, on two pages the first coding never cited, \eIcode{EQ-CLAUDE-08} and \eIcode{EQ-DEVIN-08}, and a later evidence audit refuted a fifth on \eIcode{CURSOR-DOC-06}. Three claims survived. One of the three was materially qualified by \eIcode{EQ-CURSOR-06}, so the A.3 cell records that what Cursor does not document is the direction of authorship rather than attribution as such. All three are carried in the map as NOT \nohyphens{DOCUMENTED} cells with their page sets. The four cells that carried the marker in the version the second coder worked from are documented now, at A.7, A.8 and A.9 for Claude Code and at A.9 for Devin. The four-step protocol those claims were recorded and re-tested under is deposited as \eIcode{absence\_survival.py}, and the claim-by-claim correction history of all eight, with each page set and outcome, with the evidence map of Appendix A. Where a page inside a searched set comes near a claim without meeting it, the register names it rather than passing over it: at Devin's plan approval it now names the Security Swarm threat-model pause (\eIcode{EQ-DEVIN-06}) beside the environment blueprint already recorded there, rejected on the same ground, the pause gating an investigation rather than an assigned code change; and it names the two Cursor pages a reader testing the repository-permission and deployment claims reaches first (\eIcode{CURSOR-DOC-13}, \eIcode{EQ-CURSOR-16}).

\subsection{Coding and trustworthiness}\label{sec:ms3.4}

The procedure ran in seven steps, from corpus definition and retrieval through to scoring with disagreement analysis and reconciliation. The first three are reproducible from the manifest by the deposited scripts without any model, the deposited appendix tabulates each step against the instrument that performed it, and only the mechanism classification on the two axes carries a coefficient.

\textbf{Who did what.} The first author defined the corpus, wrote the coding frame, and made every first-pass coding decision, alone. The blind second coding was performed by a person independent of this study and of the classification, working from the archived documents and written instructions that withheld the original labels and required that the paper not be read first. AI assistance to the first author is scoped to the corpus, retrieval and extraction steps and the deposited scripts; no model made any of his coding decisions. A third coder, the second author, who joined this paper after coding, later re-coded three Table 3 mechanisms, thirty-two labels in the map and all twelve NOT \nohyphens{DOCUMENTED} cells from the archived documents and the written rules before reading any part of the manuscript, with an AI assistant sweeping the page sets and drafting labels under the rule that every returned label was his own decision.

The reference labels were fixed in code before the sheet was issued: \eIcode{coding/\eIbrk{}recode.py} holds the first coder's decision on all twenty-five rows and regenerates the sheet together with the separate gold-label file the second coder was instructed not to open. The second coding was a census of the sheet as issued, not a sample: thirteen coding units in twenty-five decisions, thirteen on \eIcode{compels} and twelve on \eIcode{actor}, the difference being the Claude Code merge-gate row, an absence claim with no acting party. Six Table 3 rows post-date it, R4, R5, R14 and R16 to R18, as do Figure~\ref{fig:grid}'s four bottom-right occupants and Appendix A's thirty-two labels in the map. Section~\ref{sec:Threats} reports the coefficients.

\textbf{What the design supports.} Credibility rests on the blind second coding and on every claim being keyed to an archive item identifier whose body is stored byte-exact; transferability on the deposited codebook and on a corpus bounded by a selection rule and a retrieval date; dependability on retrieval and extraction running as deposited scripts with no model in the path. The map, the three patterns and the two cases were coded and synthesised by one analyst who also holds the thesis of Section~\ref{sec:ms1.2}, and Section~\ref{sec:Threats} names the surfaces carrying no second reading.

\subsection{The map, its availability, and the standing of the materials}\label{sec:ms3.5}

Nine workflow events form the columns: task assignment, plan production and approval, change authorship, pull request creation, check execution, review, approval, merge, and deployment. Each was selected on one criterion, that at least one collected document describes it as a discrete step at which an actor acts. The rows are the five dimensions of Section~\ref{sec:ms3.2}. The dimensions coded vary by event under rule C-2, and the deposited appendix tabulates that judgement for all forty-five event-by-dimension pairs, twelve of which take \emph{Inapplicable}. \emph{Carried in Table 3} occurs once, at A.7, where the check \emph{is} the event, so that the verification mechanisms the cell contains go on the two axes into Table 3. Record is inapplicable at A.5 because the workflow run \emph{is} the check's execution and not a separate artifact; where a run is a record it records the event that started it, at A.1.

\textbf{Availability.} The source collection, the archived documents, the extraction and absence records, the frame, the full evidence map of Appendix A (\eIcode{appendix\_A\_evidence\_map.md}), the completed second coding, and the scripts that produced all of it are deposited at \url{https://doi.org/10.5281/zenodo.21965181}~\cite{Farrag2026data}, the concept DOI, which resolves to the latest version. Every quotation in Appendix A can be checked against the archived body it cites, deposited byte-exact with its retrieval header. The coding sheets and the commands that recompute the coefficients of Section~\ref{sec:Threats} with their intervals and exact tests are deposited with it, and Zenodo publishes a checksum for the archive file itself. \eIcode{consistency\_check.py}, run with the manuscript source as its argument, checks every count this paper asserts about the collection against \eIcode{archive/\eIbrk{}retrieval\_index.csv} and resolves every archive item identifier the text cites: nineteen checks passed, \eIcode{MISMATCHES: 0}, exit code 0, and sixty-nine identifiers cited in the printed text, all sixty-nine in the manifest and all sixty-nine returning HTTP 200. Two evidence citations the second coder found unsupported in his issued sheet are corrected in the deposit; both are display cells, so neither correction moves any figure in Section~\ref{sec:Threats}.

\textbf{No human subject, and no personal data.} The 121 items are documents their publishers serve to any reader. No individual is the subject of any observation, quotation or characterisation. Four tools and seven providers are named throughout because a claim about particular published instruments cannot be checked against an anonymised document, and everything this study produced is released under CC BY 4.0.

\textbf{Two people other than the first author took part.} Both disagreements of the blind second coding are reported in Section~\ref{sec:Threats} with the reasoning on each side, so that neither reads as an assessment of the coder, who is named nowhere in this paper or the deposit. The third coder, the second author, coded three sheets blind from the archive before reading the manuscript and then joined the paper as co-author; it is that sequence, not the absence of a stake, that makes the coding reportable.

\textbf{The authors' position relative to what was studied.} Both authors used Claude, a product of Anthropic, in preparing this work, and the declaration printed with the end matter states where. Anthropic is one of the seven providers whose instruments are coded here, and Claude Code one of the four tools whose documentation is read. That position enters the work through the frame, which is fixed in advance and published in the deposit, so a reader who rejects it can re-code the same archived bodies under another. A model that drafts prose can also fabricate a quotation, a count or a citation, and the answer to that risk is the keying of every claim to an archive item identifier whose body is stored byte-exact and which \eIcode{consistency\_check.py} resolves. No language model made a coding decision of the first author's, and every coding decision of the second author's was checked and adopted by him.

\section{Two documented cases}\label{sec:Cases}

Two providers state the disagreement inside their own material. Neither case compares providers and neither is evidence about conduct.

\subsection{A responsibility to evaluate the use of any suggestion, and a product that performs the evaluation}\label{sec:ms6.1}

Cursor's terms of service, last updated 13 August 2026, contain two clauses, printed verbatim from the archived instrument (\eIcode{POL-CURSOR-01}).

Clause 1.4, Limitations for Suggestions, ends: ``You agree that you are responsible for evaluating, and bearing all risks associated with, the use of any Suggestions, including any reliance on the accuracy, completeness, or usefulness of Suggestions.'' (\eIcode{POL-CURSOR-01})

Clause 1.7, Auto-Code Execution, reads in full: ``The Service may include a feature that automatically executes code Suggestions without manual review or confirmation, and will be clearly labeled accordingly. By enabling this feature, you acknowledge and agree that you are assuming all risks associated with the execution of automatically generated code, including without limitation system outages, software defects, data loss, and security vulnerabilities. YOU ARE SOLELY RESPONSIBLE FOR ANY IMPACT RESULTING FROM USE OF THIS FEATURE, INCLUDING ENSURING APPROPRIATE SAFEGUARDS, TESTING, AND MONITORING ARE IN PLACE.'' (\eIcode{POL-CURSOR-01})

Read as a pair, the two are a general rule and a specific opt-in. Clause 1.7 does not remove the duty of 1.4. It adds, for that mode, a system-level duty to keep safeguards, testing and monitoring in place, its affirmative words sitting inside an \emph{including} gloss attached to an all-capitals assumption of risk. It is the one clause in the collection that names autonomous execution.

The pattern is not confined to one provider. GitHub's responsible-use documentation calls the permission prompt a key safety mechanism for local agentic execution, and then describes an autopilot mode in which the user grants full permissions so that a task completes without approving activity along the way (\eIcode{GH-AGENT-06}): a key safety mechanism and a mode that removes it, in one document, from the provider whose terms carry the explicit duty to review, test and validate. That is a command-line feature rather than a control on the cloud agent in a repository.

Against this, Cursor's documentation. Thirty-two pages were checked, covering agent modes and overview, review guidance, permissions, run modes, cloud agents, automations, the GitHub integration, the review product and the merge queue: the whole archived Cursor collection, twenty-six of them distinct bodies, read in full twice and swept by the deposited patterns (Section~\ref{sec:ms3.3}). No page states a general requirement that a human approve, review or confirm before an agent's changes are merged. Two narrower ones are stated, both on the page this case rests on and both governing whether the agent approves rather than whether the merge proceeds: the product will not approve a pull request where an enabled Bugbot or security review reports findings that need human review, and, where the pull request alters an approval policy, routing file, routed policy file or reviewer-specific policy file, it uses the base-branch version or, failing that, requires human review (\eIcode{CURSOR-DOC-08}). Neither reaches a pull request that changes no policy file and draws no reviewer finding, which is the case the product is offered for.

What stands together in this provider's own material is therefore not clause 1.4 against clause 1.7 but clause 1.4 against a product. The PR Routing \& Approval product and the automations that grant an agent approval rights (\eIcode{CURSOR-DOC-07}, \eIcode{CURSOR-DOC-08}) perform the evaluation clause 1.4 makes the user's responsibility, against criteria the user configured once, and record the result as an approval. No clause addresses that mode, and an agent approving a pull request is not, on the ordinary reading, a feature that executes code suggestions.

On the ordinary contracts reading, clause 1.4's responsibility for evaluating survives the user delegating the evaluating to anything, the provider's own classifier included, so on that reading there is no contradiction. The point here does not depend on responsibility having moved. It is that the artifact the platform emits is the same whether a person or the classifier formed the judgement, so the allocation clause 1.4 makes cannot be checked against the record.

\subsection{A product that describes an agent and terms that describe a return value}\label{sec:ms6.2}

GitHub's product documentation describes an autonomous agent: it authors commits, opens pull requests, and cannot approve or merge a pull request, so that the draft pull requests it creates must be reviewed and merged by a human (\eIcode{GH-AGENT-04}). GitHub's terms describe something else.

Section J of the terms of service (\eIcode{POL-GH-01}), effective April 2026, governs AI features and makes the user responsible for reviewing, testing, and validating any output before use. Its vocabulary is fixed in the instrument's definitions, where an AI feature generates output, input is content provided to it, and output is what it generates. Section J contains no instance of agent, agentic, autonomous, automated, delegated, on your behalf, without human, unattended, or take action. The only provision in the instrument that addresses acting through a non-human party is in the account requirements, where machine accounts are set up by an individual human who accepts the terms on the account's behalf and leave the owner ultimately responsible for the machine's actions. That provision is the contractual construct under which agentic work currently sits.

The terms governing enterprise customers were replaced in March 2026 (\eIcode{POL-GH-02}), and the replacement introduces the word agent, four times, in a single sense: the customer is solely responsible for any application or agent they create, and for third-party agent templates they install. The agent contemplated is one the customer builds; the agent that opens pull requests in the customer's repository is still described only through its output. That document contains no duty to verify, review, validate, or test, and neither does the GitHub Customer Agreement's General Terms. Searching all three, together with the Enterprise Cloud product-specific terms, for the twenty-one terms expressing a duty to check output before use returns nothing in the Generative AI Services Terms and one term in each of the other two, neither imposing such a duty, against nine of the twenty-one in the terms of service governing individual users.

The single clause in the enterprise document that allocates responsibility is headed Shared Responsibility: ``While GitHub is responsible for our commitments in this document, you are solely responsible for any application or agent you create using (or for use with) Generative AI Services'' (\eIcode{POL-GH-02}). Shared responsibility is settled vocabulary in cloud contracting, a division between provider-secured infrastructure and customer-secured configuration, and has never connoted symmetry. The division in what the instruments expressly allocate is therefore this: an individual user carries an explicit duty to review, test and validate output before use, while a volume licence customer is subject to none in the three documents GitHub serves directly, \eIcode{POL-GH-02}, \eIcode{POL-GH-04} and \eIcode{POL-GH-07}, all three searched.

Two qualifications, and the second is a finding. \textbf{Absence of a term is not absence of a duty}: duties arise from background law whether or not an instrument recites them, so what the search establishes is a fact about drafting. \textbf{The obvious explanation is not sufficient on the documents searched.} On the drafting-convention account, consumer terms carry express duties because consumer protection requires salient disclosure, and negotiated business instruments omit them because background commercial law fills the gap. The collection contains another provider with both a consumer and a commercial instrument, and the same search applies to it.

\begin{figure}[!htbp]
\centering
\setlength{\fboxsep}{0pt}
\setlength{\fboxrule}{0.4pt}
\newcommand{\eIvon}{\fcolorbox[gray]{0.30}{0.22}{\makebox[2.6mm][c]{%
  \rule[-0.3ex]{0pt}{1.7ex}}}}
\newcommand{\eIvoff}{\fcolorbox[gray]{0.78}{1}{\makebox[2.6mm][c]{%
  \rule[-0.3ex]{0pt}{1.7ex}}}}
\newcommand{\eIvh}[1]{\makebox[2.6mm][l]{\rotatebox[origin=lB]{60}{\tiny #1}}}
\setlength{\tabcolsep}{0.4mm}
\renewcommand{\arraystretch}{1.05}
\begin{tabular}{@{}p{48mm}@{\hspace{1.6mm}}*{21}{c}@{\hspace{2mm}}r@{}}
 & \eIvh{verify} & \eIvh{verifying} & \eIvh{verification} & \eIvh{validate} & \eIvh{validating} & \eIvh{validation} & \eIvh{review} & \eIvh{reviewing} & \eIvh{test} & \eIvh{testing} & \eIvh{evaluate} & \eIvh{evaluating} & \eIvh{evaluation} & \eIvh{inspect} & \eIvh{check~the} & \eIvh{before~use} & \eIvh{before~using} & \eIvh{before~relying} & \eIvh{accuracy} & \eIvh{human~oversight} & \eIvh{human~review} & \eIvh{of 21} \\[0.8mm]
\raggedright\scriptsize \textbf{Sourcegraph AI Terms of Use} & \eIvoff & \eIvoff & \eIvoff & \eIvoff & \eIvon & \eIvoff & \eIvon & \eIvon & \eIvoff & \eIvoff & \eIvoff & \eIvoff & \eIvoff & \eIvoff & \eIvoff & \eIvon & \eIvoff & \eIvoff & \eIvon & \eIvoff & \eIvoff & {\scriptsize\textbf{5}} \tabularnewline
\raggedright\scriptsize \textbf{GitHub Generative AI Services Terms (volume)} & \eIvoff & \eIvoff & \eIvoff & \eIvoff & \eIvoff & \eIvoff & \eIvoff & \eIvoff & \eIvoff & \eIvoff & \eIvoff & \eIvoff & \eIvoff & \eIvoff & \eIvoff & \eIvoff & \eIvoff & \eIvoff & \eIvoff & \eIvoff & \eIvoff & {\scriptsize\textbf{0}} \tabularnewline
\noalign{\vspace{0.6mm}\hrule\vspace{0.9mm}}
\raggedright\scriptsize GitHub Terms of Service (individual) & \eIvoff & \eIvoff & \eIvoff & \eIvoff & \eIvon & \eIvoff & \eIvon & \eIvon & \eIvon & \eIvon & \eIvoff & \eIvon & \eIvon & \eIvoff & \eIvoff & \eIvon & \eIvoff & \eIvoff & \eIvon & \eIvoff & \eIvoff & {\scriptsize 9} \tabularnewline
\raggedright\scriptsize Cursor Terms of Service & \eIvoff & \eIvoff & \eIvoff & \eIvoff & \eIvoff & \eIvoff & \eIvon & \eIvoff & \eIvon & \eIvon & \eIvoff & \eIvon & \eIvon & \eIvoff & \eIvon & \eIvoff & \eIvoff & \eIvoff & \eIvon & \eIvoff & \eIvoff & {\scriptsize 7} \tabularnewline
\raggedright\scriptsize Anthropic Consumer Terms & \eIvoff & \eIvoff & \eIvoff & \eIvoff & \eIvoff & \eIvoff & \eIvon & \eIvoff & \eIvon & \eIvoff & \eIvon & \eIvoff & \eIvon & \eIvoff & \eIvoff & \eIvoff & \eIvoff & \eIvoff & \eIvon & \eIvoff & \eIvon & {\scriptsize 6} \tabularnewline
\raggedright\scriptsize Anthropic Commercial Terms & \eIvon & \eIvoff & \eIvoff & \eIvoff & \eIvoff & \eIvoff & \eIvon & \eIvoff & \eIvoff & \eIvoff & \eIvon & \eIvoff & \eIvoff & \eIvoff & \eIvoff & \eIvoff & \eIvon & \eIvoff & \eIvon & \eIvoff & \eIvon & {\scriptsize 6} \tabularnewline
\raggedright\scriptsize JetBrains AI Service terms & \eIvon & \eIvoff & \eIvoff & \eIvoff & \eIvoff & \eIvoff & \eIvon & \eIvoff & \eIvoff & \eIvoff & \eIvoff & \eIvon & \eIvon & \eIvoff & \eIvon & \eIvoff & \eIvoff & \eIvoff & \eIvon & \eIvoff & \eIvoff & {\scriptsize 6} \tabularnewline
\raggedright\scriptsize GitHub Terms for Additional Products and Features (volume) & \eIvoff & \eIvoff & \eIvoff & \eIvoff & \eIvoff & \eIvoff & \eIvoff & \eIvoff & \eIvon & \eIvon & \eIvoff & \eIvoff & \eIvon & \eIvoff & \eIvoff & \eIvoff & \eIvoff & \eIvoff & \eIvoff & \eIvoff & \eIvoff & {\scriptsize 3} \tabularnewline
\raggedright\scriptsize Replit terms & \eIvoff & \eIvoff & \eIvoff & \eIvoff & \eIvoff & \eIvoff & \eIvon & \eIvoff & \eIvoff & \eIvoff & \eIvoff & \eIvoff & \eIvoff & \eIvoff & \eIvoff & \eIvoff & \eIvoff & \eIvoff & \eIvon & \eIvoff & \eIvoff & {\scriptsize 2} \tabularnewline
\raggedright\scriptsize Sourcegraph Terms of Service & \eIvoff & \eIvoff & \eIvoff & \eIvoff & \eIvoff & \eIvoff & \eIvon & \eIvoff & \eIvon & \eIvoff & \eIvoff & \eIvoff & \eIvoff & \eIvoff & \eIvoff & \eIvoff & \eIvoff & \eIvoff & \eIvoff & \eIvoff & \eIvoff & {\scriptsize 2} \tabularnewline
\raggedright\scriptsize Cognition terms & \eIvoff & \eIvoff & \eIvoff & \eIvoff & \eIvoff & \eIvoff & \eIvoff & \eIvoff & \eIvoff & \eIvoff & \eIvoff & \eIvoff & \eIvoff & \eIvoff & \eIvoff & \eIvoff & \eIvoff & \eIvoff & \eIvon & \eIvoff & \eIvoff & {\scriptsize 1} \tabularnewline
\raggedright\scriptsize GitHub Customer Agreement, General Terms (volume) & \eIvoff & \eIvoff & \eIvoff & \eIvoff & \eIvoff & \eIvoff & \eIvoff & \eIvoff & \eIvoff & \eIvoff & \eIvoff & \eIvoff & \eIvon & \eIvoff & \eIvoff & \eIvoff & \eIvoff & \eIvoff & \eIvoff & \eIvoff & \eIvoff & {\scriptsize 1} \tabularnewline
\raggedright\scriptsize GitHub Enterprise Cloud product terms (volume) & \eIvoff & \eIvoff & \eIvoff & \eIvoff & \eIvoff & \eIvoff & \eIvoff & \eIvoff & \eIvoff & \eIvoff & \eIvoff & \eIvoff & \eIvon & \eIvoff & \eIvoff & \eIvoff & \eIvoff & \eIvoff & \eIvoff & \eIvoff & \eIvoff & {\scriptsize 1} \tabularnewline
\raggedright\scriptsize Sourcegraph Cody Enterprise Terms of Use & \eIvoff & \eIvoff & \eIvoff & \eIvoff & \eIvoff & \eIvoff & \eIvoff & \eIvoff & \eIvoff & \eIvoff & \eIvoff & \eIvoff & \eIvoff & \eIvoff & \eIvoff & \eIvoff & \eIvoff & \eIvoff & \eIvon & \eIvoff & \eIvoff & {\scriptsize 1} \tabularnewline
\end{tabular}

\vspace{1.2mm}
{\scriptsize\mbox{\eIvon\ the term occurs in the archived body}\quad\mbox{\eIvoff\ it does not}}
\caption{Express verification-duty vocabulary, present or absent: the
twenty-one terms across the fourteen of the eighteen collected policy
documents that are instruments, Section~\ref{sec:ms4.4} naming the four
that are not. The right-hand column counts the row's filled cells, out of
twenty-one, and seven terms are filled in no instrument. Above the rule
are the two AI-specific supplements Section~\ref{sec:ms6.2} compares; the
twelve below are ordered by that count, ties by provider. The word counts,
occurrence totals, term list and matching rule are deposited with the
evidence map of \ref{sec:A}.}
\label{fig:vocabgrid}
\end{figure}

A filled cell is a word, and only reading it in context settles whether it states a duty. Three cells are filled in the row for GitHub's Terms for Additional Products and Features, and none of the occurrences behind them is a verification duty: reading every one in context returns a development-and-testing licence scope, benchmark testing, and the definition of previews (\eIcode{POL-GH-05}).

That other provider is Anthropic, whose consumer and commercial instruments are filled at the same number, six of the twenty-one each, where GitHub's individual-to-volume boundary carries the fall from nine. The drafting-convention account therefore holds at one of the two providers that publish both and not at the other, which is not enough to explain the asymmetry away.

A sharper comparison does not depend on that boundary. Three providers publish an instrument addressed specifically to AI services on its own title, and two are short enough for a like-for-like comparison: the two rows above the rule in Figure~\ref{fig:vocabgrid}. GitHub's Generative AI Services Terms runs to 951 words and contains none of the twenty-one terms. Sourcegraph's AI Terms of Use runs to 647 words and contains five of them, including the sentence ``You are solely responsible for reviewing and validating any Outputs before use'' (\eIcode{POL-SG-04}). Both are supplements incorporated into a larger agreement, of comparable length, written for the same product class and addressed to business customers, so length, the sophistication of the counterparty and the supplement's place in a hierarchy do not explain the difference. One explanation stays untested: the documents GitHub's agreement routes to were not retrieved (Section~\ref{sec:Threats}). The comparison is not a rate. It shows what an AI-specific supplement can state, not how often one does.

An earlier archive of Sourcegraph's terms captured a 266-word index page and reported the provider at zero; retrieving the instruments it points to reversed the finding (Section~\ref{sec:ms3.2}). The Terms for Additional Products and Features effective April 2026 (\eIcode{POL-GH-05}), incorporated by Section I of the terms of service (\eIcode{POL-GH-01}), direct Copilot Business and Copilot Enterprise users to a product-specific terms document whose first line states ``These terms have been deprecated effective 5 March 2026'' (\eIcode{POL-GH-06}). A current document points at a superseded one.

\section{The map}\label{sec:Results}

Three results come before the detail. \textbf{First, the terms give no event-level account.} At five of the nine events no collected instrument names the event in its workflow sense, so the terms layer offers no workflow-level statement there to compare the documentation against (Section~\ref{sec:ms4.4}). \textbf{Second, verification has two axes rather than one:} whether a mechanism compels a check, and whether a person performs it. One provider's documentation describes a product that performs the evaluation of output its own terms make the user's responsibility, and records the result as an approval (Section~\ref{sec:ms4.2}). \textbf{Third, the artifacts record actions and identities and not the performance of a verification duty} (Section~\ref{sec:ms4.3}, which answers RQ3).

The map is how those results were found rather than the finding itself. Each subsection below takes one or more of the five dimensions across the nine events, records under RQ2 what each layer states there, and puts RQ1 to it. Figure~\ref{fig:eventmap} summarises the map, which the deposited Appendix A gives in full for all four tools.

\begin{figure}[!htbp]
\centering
\newcommand{\eIrot}[1]{\makebox[7.5mm][c]{\rotatebox[origin=lB]{55}{\footnotesize #1}}}
\newcommand{\eIsw}[3]{\fcolorbox[gray]{#1}{#1}{\makebox[7.5mm][c]{%
  \rule[-0.65ex]{0pt}{2.9ex}\textcolor{#2}{#3}}}}
\newcommand{\eIfour}{\eIsw{0.24}{white}{4}}
\newcommand{\eIthree}{\eIsw{0.50}{white}{3}}
\newcommand{\eItwo}{\eIsw{0.72}{black}{2}}
\newcommand{\eIone}{\eIsw{0.88}{black}{1}}
\newcommand{\eIper}{\eIsw{0.94}{black}{P}}
\newcommand{\eIna}{\fcolorbox[gray]{0.80}{1}{\makebox[7.5mm][c]{%
  \rule[-0.65ex]{0pt}{2.9ex}\textcolor[gray]{0.55}{--}}}}
\newcommand{\eIcar}{\fcolorbox[gray]{0.80}{1}{\makebox[7.5mm][c]{%
  \rule[-0.65ex]{0pt}{2.9ex}\textcolor[gray]{0.35}{\tiny T}}}}
\footnotesize
\setlength{\tabcolsep}{1.1pt}
\renewcommand{\arraystretch}{1.1}
\begin{tabular}{@{}r@{\hspace{2.5mm}}*{9}{c}@{}}
 & \eIrot{A.1 Task assignment} & \eIrot{A.2 Plan approval}
 & \eIrot{A.3 Change authorship} & \eIrot{A.4 Pull request creation}
 & \eIrot{A.5 Check execution} & \eIrot{A.6 Review} & \eIrot{A.7 Approval}
 & \eIrot{A.8 Merge} & \eIrot{A.9 Deployment} \\[1.5mm]
Authority    & \eIfour & \eIthree & \eIna    & \eIfour & \eIna   & \eIna   & \eIfour  & \eIfour  & \eIthree \\
Execution    & \eIfour & \eIna    & \eIthree & \eIfour & \eIna   & \eIna   & \eIfour  & \eIna    & \eIna   \\
Verification & \eIna   & \eIthree & \eIna    & \eIna   & \eIfour & \eIfour & \eIcar   & \eIfour  & \eIthree \\
Consequence  & \eIper  & \eIper   & \eIper   & \eIper  & \eIper  & \eIper  & \eIper   & \eIper   & \eIper  \\
Record       & \eIfour & \eIthree & \eIfour  & \eIfour & \eIna   & \eIfour & \eIfour  & \eIfour  & \eIone  \\
\end{tabular}

\vspace{1.2mm}
{\scriptsize\mbox{\eIfour\,\eIthree\,\eItwo\,\eIone\ tools documenting an
answer, of four}\quad\mbox{\eIper\ consequence, once per provider}\quad
\mbox{\eIna\ inapplicable}\quad\mbox{\eIcar\ carried into Table~\ref{tab:3}}}
\caption{The Accountability Event Map at a glance: nine workflow events across,
the five accountability dimensions down, over the four tools of
\ref{sec:A}. A numeral is how many of those four tools the map records
a documented answer for in that cell, out of four, and the shading runs with it.
A numeral is documentation coverage and not sufficiency: it counts the tools
that document an answer at that cell, not the tools whose artifact would answer
a later question about who judged the change.
A dash is a dimension judged inapplicable at that event under rule C-2, whose
matrix is the one Section~\ref{sec:ms3.5} deposits. T is the one pair that matrix
records as carried into Table~\ref{tab:3} rather than inapplicable: verification at
approval, where the check is the event. P is consequence, coded once per
provider and at no event at all under rule C-1.}
\label{fig:eventmap}
\end{figure}

The Record row is full but for two events: at plan approval one of the four tools documents no artifact, and at deployment, its emptiest column, three of the four do not, though two of the three now document a gate (Section~\ref{sec:ms4.2}). Ninety-two cells stand behind Figure~\ref{fig:eventmap}, the twenty-three applicable event-by-dimension pairs across four tools; nine are undocumented, and eight of the nine fall in those two columns; eight of the nine carry the marker itself and the ninth is the declared label \eIcode{compels = Not documented} at deployment, which Appendix A reports separately. GitHub Copilot is documented at every cell.

\subsection{Authority and execution}\label{sec:ms4.1}

Authority is the most consistently documented dimension, and the most constrained. GitHub restricts the cloud agent to users with repository write access, never presents it comments from users without that access, and its automations ignore events from untrusted users by default (\eIcode{GH-AGENT-06}, \eIcode{GH-AGENT-04}). Its risks-and-mitigations page states that the agent ``cannot approve or merge a pull request'', so that ``draft pull requests created by Copilot cloud agent must be reviewed and merged by a human'' (\eIcode{GH-AGENT-04}). Where approvals are required, the assigning user's approval does not count toward the required number and another reviewer must approve before merge (\eIcode{GH-AGENT-07}), which bites only where the rule Table 3 records at R2 has been configured. The agent works on one branch at a time and opens exactly one pull request per task (\eIcode{GH-AGENT-01}).

One constraint is stated with its rationale. The platform prevents the user who asked the cloud agent to create a pull request from approving it, on the stated ground that this maintains the expected controls in the required-approvals rule and branch protection (\eIcode{GH-AGENT-04}), and the same reasoning covers automations generally, whose pull requests are attributed to their creator, who cannot approve them either (\eIcode{EQ-GITHUB-03}). It is the collection's clearest statement of a deliberately encoded accountability boundary.

The same vendor names an affordance the constraint does not account for. Its Copilot app page records that users can inspect work before submitting reviews, enabling agent merge, or merging pull requests (\eIcode{GH-AGENT-06}), and no archived page explains what enabling agent merge is. One page outside the archive, read on 10 September 2026 without the retrieval discipline of Section~\ref{sec:ms3.3}, does: it prompts the workspace's Copilot session to read the pull request, fix what is blocking it, and merge it as soon as GitHub allows~\cite{GitHubAgentMerge2026}. Together they describe an agent that cannot approve or merge the pull request it authored, and an agent session that merges one on a user's instruction once the platform's own rules permit.

Execution divides differently. Copilot's commits are authored by the agent with the assigning human as co-author, signed, shown as verified, each commit message linking to the agent session logs (\eIcode{GH-AGENT-05}). Devin documents chat-agent commits as the Devin bot, and comments a user writes through the review interface under that user's own GitHub identity (\eIcode{DEVIN-02}). Claude Code documents replies posted to review threads under the human's GitHub account and username, distinguished only by a label in the reply body (\eIcode{CC-DOC-06}). The direction of attribution is inverted between providers, and both directions are current, documented, and applied to the same artifact type in the same platform.

\footnotesize
\begin{longtable}{>{\raggedright\arraybackslash}p{1.85cm}>{\raggedright\arraybackslash}p{2.35cm}>{\raggedright\arraybackslash}p{1.69cm}>{\raggedright\arraybackslash}p{2.33cm}>{\raggedright\arraybackslash}p{2.10cm}>{\raggedright\arraybackslash}p{2.02cm}}
\caption{One tool's slice of the map, GitHub Copilot cloud agent. A dash marks a pair the matrix of Section~\ref{sec:ms3.5} judges inapplicable at that event; at approval the verification mechanisms are carried into Table 3 (R3, R15). Consequence is coded once per provider under rule C-1.}\label{tab:1} \\
\toprule
\textbf{Event} & \textbf{Authority} & \textbf{Execution and recorded identity} & \textbf{Verification} & \textbf{Consequence} & \textbf{Record} \\
\midrule
\endfirsthead
\multicolumn{6}{@{}l}{\footnotesize\itshape Table~\thetable, continued.}\\*
\toprule
\textbf{Event} & \textbf{Authority} & \textbf{Execution and recorded identity} & \textbf{Verification} & \textbf{Consequence} & \textbf{Record} \\
\midrule
\endhead
Task assignment & Write access only & Human & --- & User (per instrument) & Issue assignment \\
Plan approval & Agent produces the plan & --- & Reviewable, not gated & User (per instrument) & Plan visible in session \\
Change authorship & --- & Agent as author, human as co-author & --- & User (per instrument) & Signed commit, session log link \\
Pull request creation & Agent; draft, and some entry points open one automatically & Agent & --- & User (per instrument) & Pull request \\
Check execution & --- & --- & Compelled: a human approves before workflows run, on by default (R1) & User (per instrument) & --- \\
Review & --- & --- & Advised & User (per instrument) & Comment, if submitted \\
Approval & Not the assigning user; not the agent & A different human & Carried in Table 3 (R3, R15) & User (per instrument) & Review submission event \\
Merge & Human only & --- & Compelled: the agent cannot approve or merge its own pull request (R15) & User (per instrument) & Merge event \\
Deployment & Named reviewers & --- & Compelled where configured (R6) & User (per instrument) & Environment approval \\
\bottomrule
\end{longtable}
\normalsize

The verification column's agent-performed cell, and the plan-approval gate two other tools document at row R7, appear only when the map is extended to all four tools, in the deposited map of Appendix A.

\subsection{Verification}\label{sec:ms4.2}

This study first read verification as governance frameworks read oversight, on one scale: a mechanism withholds progress, or the documentation advises, or nothing is said. That scale conflates two independent questions. One is whether a mechanism compels verification before the work proceeds; the other is whether the party performing it is a person.

\begin{figure}[!htbp]
\centering
\newcommand{\eIgh}[1]{\makebox[58mm][c]{\footnotesize\textbf{#1}}}
\setlength{\tabcolsep}{1.2mm}
\hyphenpenalty=10000 \exhyphenpenalty=10000
\begin{tabular}{@{}p{20mm}|p{58mm}|p{58mm}|@{}}
\multicolumn{1}{@{}l}{} & \multicolumn{1}{c}{\eIgh{Performed by a person}}
 & \multicolumn{1}{c}{\eIgh{Performed by a non-person}} \\
\cline{2-3}
\raggedright\footnotesize\textbf{Compelled}
 &\raggedright\scriptsize\vspace{0.4mm}
  Required approvals\newline
  Approval by non-pusher\newline
  Workflow execution approval on agent pull requests\newline
  Plan approval before edits\newline
  Auto-mode classifier at merge, approval and deployment, Claude Code
  (\eIcode{EQ-CLAUDE-08})\newline
  Approve/deny prompt before a hosted deployment, Devin
  (\eIcode{EQ-DEVIN-08})\vspace{0.8mm}
 &\raggedright\scriptsize\vspace{0.4mm}
  PR Routing and Approval below a configured risk threshold, Cursor\newline
  Merge queue admission, Cursor\newline
  Required status checks before merging, performed by a check or analyser
  (\eIcode{GH-CTRL-01})\dag\vspace{0.8mm} \tabularnewline
\cline{2-3}
\raggedright\footnotesize\textbf{Not compelled}
 &\raggedright\scriptsize\vspace{0.4mm}
  Advisory instruction to review before merging\vspace{0.8mm}
 &\raggedright\scriptsize\vspace{0.4mm}
  Bugbot, which reviews every pull request and flags bugs, security issues and
  code quality problems (\eIcode{CURSOR-DOC-16})\newline
  Security Reviewer, which checks pull requests before they merge on Git-based
  triggers (\eIcode{EQ-CURSOR-02})\newline
  Claude Code auto-fix, which responds automatically to CI failures and review
  comments (\eIcode{CC-DOC-06})\newline
  Devin Review, which creates a commit status check on the pull request for
  each review (\eIcode{DEVIN-02})\vspace{0.8mm} \tabularnewline
\cline{2-3}
\end{tabular}
\caption{Verification on two axes: whether the mechanism withholds progress
until the check occurs, and whether the party performing it is a person. The
column records only whether the performer is a person; the frame's
\eIcode{Agent} label covers every non-person, and Section~\ref{sec:ms5.1}
states what the axis does not record. The grid holds fifteen mechanisms,
eleven of them rows of Table~\ref{tab:3} and four further instances it does
not tabulate, so these fifteen are not the fifteen compelling rows of
Table~\ref{tab:3}. A dagger marks a non-person that is
not an AI coding agent. Seven of the fifteen carry the agreement statistic of
Section~\ref{sec:Threats}; R14, R16, R17 and R18 were read by the third coder
and not the second, and the four in the bottom-right cell were put to neither.}
\label{fig:grid}
\end{figure}

At the retrieval date one vendor occupied the top-right cell, which a one-scale reading cannot express, beside required status checks (R14), a compelled check performed by neither a person nor a coding agent; Section~\ref{sec:ms5.1} states why the two occupants differ. Each of the four occupants of the bottom right, Bugbot (\eIcode{CURSOR-DOC-16}), Security Reviewer (\eIcode{EQ-CURSOR-02}), Claude Code auto-fix (\eIcode{CC-DOC-06}) and Devin Review (\eIcode{DEVIN-02}), runs on every pull request or on a configured trigger rather than on request, so verification-shaped output arrives unbidden and at volume and none of it withholds anything.

\textbf{Compelled, human-performed.} Workflow execution on agent-authored pull requests is the clearest instance. Actions workflows triggered by pull requests the agent raises require approval from a user with write access, and by default are not triggered until the agent's code is reviewed and such a user clicks the approve-and-run button (\eIcode{GH-AGENT-06}, \eIcode{GH-AGENT-04}). That default conjoins two things, and only the click is a mechanism; the reading is advice, stated as such on the page giving the procedure, which also offers the option of letting workflows run without human intervention (\eIcode{EQ-GITHUB-12}). The control compels a click, not a demonstrated judgement: a user who clicks without reading passes the same gate as a user who reads.

Required approvals, when configured, compel in the same sense. Plan approval before edits, documented by two of the four tools at row R7, compels only inside a mode a person enters, which is why Table 3 records the row's basis as inferred. Two rows state a human approval requirement that no page conditions on a setting: R15, where the agent cannot approve or merge a pull request, and R18, where ``every deploy requires your explicit approval'' within the hosted-deployment feature that carries it (\eIcode{EQ-DEVIN-08}). Section~\ref{sec:ms5.3} sets those two against the thirteen that a reader turns on, turns off, or may be permitted to go around.

\textbf{Not compelled, guidance stated.} GitHub tells reviewers to check the pull request thoroughly before merging (\eIcode{GH-AGENT-07}). Cursor's guidance states that merge standards should be the same whether code was written by hand or by an agent (\eIcode{CURSOR-DOC-03}). Claude Code's action documentation says to review changes before merging (\eIcode{CC-DOC-02}).

\textbf{Compelled, agent-performed.} Cursor documents a product that occupies the approval event. PR Routing \& Approval can approve low-risk pull requests once the user's approval criteria are met, subject to a Maximum Risk Threshold that ``sets the highest risk level the agent may approve''; it withholds approval where an AI reviewer has reported findings needing human review, and the same page states that it does not replace a full code review (\eIcode{CURSOR-DOC-08}).

The threshold compels: a pull request above it is not approved. What has changed is who stands at the gate, not whether one exists, and what the record distinguishes is the actor and not the act. The audit log's field lists carry \eIcode{actor\_is\_agent}, which indicates whether the actor is an AI agent (\eIcode{EQ-GITHUB-08}), while the field list for a submitted review carries no field for the state of the review and none for what the approver assessed (\eIcode{GH-CTRL-08}, Section~\ref{sec:ms5.1}). Identity is stated at the automation layer rather than at the approval: an automation runs under the configuring user's own auth until it is promoted to Team Owned, when it runs with the team's shared automations service account (\eIcode{CURSOR-DOC-07}, \eIcode{CURSOR-DOC-08}), which no archived page says the platform would mark with \eIcode{actor\_is\_agent}. That approval blocks a merge only where required approvals, row R2, are configured, and R2's default this study records as off by inference.

\textbf{Compelled within a mode, advised outside it.} Across nineteen Claude Code documentation pages, seventeen of them distinct bodies, no page withholds the merge event as such from a person (R12), and the guidance to review before merging (\eIcode{CC-DOC-02}) withholds nothing. One page, the permission-modes reference (\eIcode{EQ-CLAUDE-08}), documents a gate inside \emph{auto mode}. There a classifier blocks by default ``Merging a pull request no human has approved'', approving the agent's own pull request, disabling CI checks, and production deploys and migrations; a blocked action is listed under a recently-denied tab from which a person retries it with a manual approval. Under rule B-1 that is \eIcode{compels = Yes}, \eIcode{actor = Human}, at merge and approval (R16) and at deployment (R17), on in the mode and off outside it. The page adds that stated boundaries can be lost if context compaction removes the message. The first coding reported the cell as nothing stated, and a third coder's exhaustive read found it (Section~\ref{sec:ms3.3}).

The assessment behind the block is a model's, the same page stating that in auto mode a second model reviews actions instead of the user. The row is nonetheless coded \eIcode{actor = Human}, because the mechanism withholds progress until a person retries the action with a manual approval, which is the party the map records. The classifier is a non-person forming a content-dependent judgement, and Section~\ref{sec:ms5.1}'s boundary separates it from the Cursor approver, because it blocks and emits no approval.

A team reading vendor documentation encounters compelling gates, advisory sentences and a product that approves pull requests, all in the same prose register and with nothing to signal which is which.

\footnotesize
\begin{longtable}{>{\raggedright\arraybackslash}p{1.52cm}>{\raggedright\arraybackslash}p{6.22cm}>{\raggedright\arraybackslash}p{3.25cm}>{\raggedright\arraybackslash}p{2.18cm}}
\caption{Cross-provider divergence. For each tool: the merge gate the vendor documents, the direction in which its documentation attributes authorship of an agent's commit, and what that provider's governing terms state about a duty to verify output before use. In the last column, \emph{None stated} and \emph{Volume licence: none} record an instrument that was collected and searched, not one that was not retrieved.}\label{tab:2} \\
\toprule
\textbf{Provider} & \textbf{Documented merge gate} & \textbf{Attribution direction} & \textbf{Verification duty in terms} \\
\midrule
\endfirsthead
\multicolumn{4}{@{}l}{\footnotesize\itshape Table~\thetable, continued.}\\*
\toprule
\textbf{Provider} & \textbf{Documented merge gate} & \textbf{Attribution direction} & \textbf{Verification duty in terms} \\
\midrule
\endhead
GitHub Copilot & Enforced for the agent's own pull request: it cannot approve or merge it; an agent-merge path on a user's instruction is documented outside the archive (Section~\ref{sec:ms4.1}) & Agent author, human co-author & Individual: explicit. Volume licence: none \\
Devin & Advisory at merge: human review recommended, merge and auto-merge available as actions in the vendor's interface; compelled at hosted deployment, where every deploy requires the user's explicit approval (\eIcode{EQ-DEVIN-08}) & Bot for agent commits, user identity for user comments & None stated \\
Cursor & Compelled, agent-performed: agent approves below a configured risk threshold; merge queue requires that the app be let push or bypass the repository's merge restrictions & Direction not documented for commits; commits signed and badged as coming from Cursor (\eIcode{EQ-CURSOR-06}); OIDC workload identity documented & Explicit, in clause 1.4, alongside clause 1.7 \\
Claude Code & Advisory outside auto mode: review changes before merging; within auto mode the classifier blocks by default the merge of a pull request no human has approved and the approval of the agent's own (\eIcode{EQ-CLAUDE-08}) & Human's own account for review replies; agent named in a co-authorship trailer on commits, disableable & Explicit in both instruments: consumer and commercial \\
\bottomrule
\end{longtable}
\normalsize

Two values in the last column record a searched instrument, not a missing one. Devin's is Cognition's, \eIcode{POL-COG-01}, in Figure~\ref{fig:vocabgrid}, whose one matching term of the twenty-one, \emph{accuracy}, sits inside an all-caps warranty disclaimer. GitHub Copilot's is the volume licence, whose three documents carry no express duty to check output where the same provider's individual terms carry one.

\footnotesize
\begin{longtable}{>{\raggedright\arraybackslash}p{2.33cm}>{\raggedright\arraybackslash}p{1.24cm}>{\raggedright\arraybackslash}p{1.38cm}>{\raggedright\arraybackslash}p{2.06cm}>{\raggedright\arraybackslash}p{1.24cm}>{\raggedright\arraybackslash}p{4.08cm}}
\caption{Verification status by mechanism. \eIcode{Compels?} and \eIcode{Acting party} are the two coded axes of Section~\ref{sec:ms3.2} and carry Section~\ref{sec:Threats}'s agreement statistic for twelve of these eighteen rows: R4, R5, R14, R16, R17 and R18 were not put to the second coder, and R15 is the unit on which both disagreements fell. \eIcode{Agent} marked with a dagger is a non-person that is not an AI coding agent. \eIcode{Default} was not second-coded; \eIcode{Basis} is \emph{stated} where the cited page states the default in terms and \emph{inferred} where the value follows from an instruction to enable, select or turn the mechanism on. Where a mechanism does not compel, both cells read n/a.}\label{tab:3} \\
\toprule
\textbf{Mechanism} & \textbf{Compels?} & \textbf{Acting party} & \textbf{Default} & \textbf{Basis} & \textbf{Source} \\
\midrule
\endfirsthead
\multicolumn{6}{@{}l}{\footnotesize\itshape Table~\thetable, continued.}\\*
\toprule
\textbf{Mechanism} & \textbf{Compels?} & \textbf{Acting party} & \textbf{Default} & \textbf{Basis} & \textbf{Source} \\
\midrule
\endhead
R1 Workflow execution approval & Yes & Human & On, disableable & stated & \eIcode{GH-AGENT-04} \\
R2 Required approvals & Yes & Human & Off & inferred & \eIcode{GH-CTRL-01} \\
R3 Approval by non-pusher & Yes & Human & Off & inferred & \eIcode{GH-CTRL-01} \\
R4 Stale approval dismissal & Yes & Human & Off & inferred & \eIcode{GH-CTRL-01} \\
R5 Code owner review & Yes & Human & Off & inferred & \eIcode{GH-CTRL-01} \\
R6 Deployment reviewers & Yes & Human & Off, admin bypass on & inferred & \eIcode{GH-CTRL-05} \\
R7 Plan approval before edits, Claude Code and Cursor & Yes & Human & On in plan mode; plan mode is opt-in & inferred & \eIcode{CC-DOC-08}, \eIcode{CURSOR-DOC-01} \\
R8 Permission prompt before system-modifying commands & Yes & Human & On, removable in autonomous modes & stated & \eIcode{CC-DOC-04}, \eIcode{CC-DOC-07}, \eIcode{EQ-CLAUDE-03} \\
R9 Agent approval below risk threshold, Cursor & Yes & \textbf{Agent} & Off & inferred & \eIcode{CURSOR-DOC-08} \\
R10 Merge queue admission, Cursor & Yes & \textbf{Agent} & Off; requires push-or-bypass rights on merge restrictions & inferred & \eIcode{CURSOR-DOC-09} \\
R11 Merge through the vendor interface, Devin & No & Human & n/a & n/a & \eIcode{DEVIN-02}, \eIcode{DEVIN-04} \\
R12 Merge gate, Claude Code, outside auto mode & No & n/a & n/a & n/a & \eIcode{CC-DOC-01}--\eIcode{CC-DOC-09}, \eIcode{EQ-CLAUDE-01}--\eIcode{EQ-CLAUDE-10} \\
R13 Reviewer instruction in documentation & No & Human & n/a & n/a & \eIcode{CC-DOC-02} \\
R14 Required status checks before merging & Yes & Agent\dag{} & Off & inferred & \eIcode{GH-CTRL-01} \\
R15 Agent cannot approve or merge its own pull request, GitHub Copilot & Yes & Human & On; no switch on the archived pages & stated & \eIcode{GH-AGENT-04}, \eIcode{GH-AGENT-06} \\
R16 Auto-mode classifier blocks merge of an unapproved pull request and approval of the agent's own, Claude Code & Yes & Human & On in terminal and VS Code on Pro, Max and Team plans, from v2.1.228; off in headless runs, on Enterprise and API keys, and in three other stated cases & stated & \eIcode{EQ-CLAUDE-08} \\
R17 Auto-mode classifier blocks production deploys and migrations, Claude Code & Yes & Human & As R16 & stated & \eIcode{EQ-CLAUDE-08} \\
R18 Approve/deny prompt before a hosted deployment, Devin & Yes & Human & On; disabled for enterprise organisations and in secure mode & stated & \eIcode{EQ-DEVIN-08} \\
\bottomrule
\end{longtable}
\normalsize

Fifteen of the eighteen mechanisms compel, on a split of \textbf{six stated and nine inferred}. The remaining three record no default at all, because a mechanism that withholds nothing has nothing to be on or off. For the six the vendor states the position in terms, among them Claude Code's auto-mode classifier at R16 and R17, on because ``On Pro, Max, and Team plans, the built-in starting mode is auto mode'' (\eIcode{EQ-CLAUDE-08}) and off in the mode Enterprise and API-key sessions start in.

The other nine are recorded as off, R7 as an opt-in mode, off until selected. For eight of them, and for R7 at Cursor, no page states a default; two are corroborated at one remove, from GitHub's statement that a reader may optionally enable additional branch protection settings and from its account of environments created implicitly by a workflow run (\eIcode{GH-CTRL-01}, \eIcode{GH-CTRL-05}), neither of which states the position for any one row. A team cannot read off these pages what it is given before it configures anything: for nine of the fifteen mechanisms that compel, the shipped position is an inference this study drew from an instruction to enable, because no vendor states it.

That split is a count over the eighteen mechanisms of Table 3 and not over everything in the corpus that carries a vendor-stated default. One archived page states a default in terms for a mechanism the table does not hold. On the automations that triage issues, ``The default level is Cautious, so only high-confidence changes apply automatically until you change it'' (\eIcode{EQ-GITHUB-15}), one of four named levels, changes below the threshold waiting in an approvals panel where a person selects Accept or Decline. Its scope is stated on the companion page: the capability covers the attributes an automation changes on an issue, and it ``does not apply to changes people make by hand, and it does not affect other actions an automation can take, such as opening pull requests or pushing code'' (\eIcode{EQ-GITHUB-01}). The one attribute it shares with the map is the assignee, at task assignment, where no verification mechanism is coded. It reaches none of the eighteen, and the split of six stated and nine inferred stands as given.

\textbf{The sentence behind every \eIcode{Default} value is deposited}, row by row with its archive item identifier, in the evidence map of Appendix A, together with the three rows whose basis is split or was corrected. Two bear on the reading here. At \textbf{R15} no archived page offers a switch, and the same vendor's app page names the agent-merge affordance defined outside the archive (\eIcode{GH-AGENT-06}, Section~\ref{sec:ms4.1}); this is also the unit on which both of the second coder's disagreements fell (Section~\ref{sec:Threats}). At \textbf{R16} and \textbf{R17} the default is stated with its exclusions, the one bearing on this paper's setting being headless sessions (\eIcode{EQ-CLAUDE-08}).

\subsection{Record}\label{sec:ms4.3}

The record dimension is the thinnest, and the thinness is the finding. Some events produce durable artifacts: a commit, a pull request, a review submission recorded in the organisation audit log, a merge. Others produce none.

The audit log's field lists carry three markers that the page enumerating the events defines none of. \eIcode{actor\_is\_agent} and \eIcode{actor\_is\_bot} appear in the field list of event after event and nowhere in prose, and \eIcode{agent\_session\_id} appears in exactly one field list, that of \eIcode{pull\_request.create}, also undefined (\eIcode{GH-CTRL-08}). Two are defined on a separate page devoted to agent events, \eIcode{actor\_is\_agent} as indicating whether the actor is an AI agent and \eIcode{agent\_session\_id} as the identifier linking to the agent session that generated the event (\eIcode{EQ-GITHUB-08}). \eIcode{actor\_is\_bot} remains undefined in either, so a reader can distinguish an agent's event from a human's but cannot tell what the platform counts as a bot.

No trailer is defined for agent authorship in any of the standardising documents collected here. GitHub's two attribution-trailer pages (\eIcode{GH-CTRL-12}, to which \eIcode{GH-CTRL-13} redirects) mention neither AI, agents nor machine authorship, and the Git project's own documentation defines the trailer mechanism without defining any attribution trailer for agents (\eIcode{GIT-TRAILER-01}). The absence of a definition has not prevented use: Claude Code's settings documentation records that its default commit attribution is a co-authorship trailer naming the model active for that session, against no definition another vendor could implement the same way, and the same page defines three keys, disposed of separately, that between them remove all attribution (\eIcode{CC-DOC-03}). Attribution here is a set of independent defaults rather than a property of the record.

This answers RQ3, on the three durability values of Section~\ref{sec:ms3.2}. \textbf{Task assignment} produces an issue assignment, a comment, an automation run and a workflow run, one at each tool (A.1), all repository- or log-durable. \textbf{Commit authorship}, \textbf{pull request creation} and \textbf{merge} produce repository-durable artifacts with no vendor-stated expiry. \textbf{Check execution} is Record-inapplicable under rule C-2; the workflow run A.1 records for GitHub Copilot and Claude Code is log-durable and deleted after thirty days while it awaits approval (\eIcode{GH-CTRL-11}). \textbf{Review} and \textbf{approval} produce an artifact only if a review is submitted, do not distinguish endorsement from comment, and are log-durable for one hundred and eighty days (\eIcode{GH-CTRL-07}, \eIcode{EQ-GITHUB-11}), so the record of who approved a change expires at that point. \textbf{Plan approval} and, at three of four tools, \textbf{deployment} produce none, and the verification the terms require produces none anywhere. The audit log's Git event class is available on GitHub Enterprise Cloud through the REST API alone, with seven-day retention (\eIcode{GH-CTRL-08}).

Cursor documents the strongest identity mechanism in the collection, the OIDC token of Appendix A.3, minted inside the agent's own VM and verifiable by any relying party against a published key set. The tokens expire five minutes after issue, and the identity is coarser than the actor, identifying the cloud agent run rather than a process inside the VM and mintable by any process that can reach the socket (\eIcode{CURSOR-DOC-11}); it attaches to outbound requests rather than to the commit. The same vendor produces a repository-facing artifact too: cloud agents sign every commit with an HSM-backed Ed25519 key, and on GitHub and GitLab those commits display a verified badge confirming the commit came from Cursor (\eIcode{EQ-CURSOR-06}). That records the producing product and not whether the human or the agent is the author, which is the correction Section~\ref{sec:ms3.3} made to the A.3 cell. Record is thin at these events not because such artifacts are hard to produce but because the ones produced answer a different question: which product or which run stood behind a request, rather than whether the party bearing the consequence formed a judgement about the change.

At one event the record is not thin but inconsistent between the layers that write it. Cursor's merge queue documentation notes that after a successful run the original pull request ``may appear as closed in GitHub instead of merged'', even though the product treats it as merged throughout (\eIcode{CURSOR-DOC-09}); the served body omits the subject of that second clause, the product name absent as published rather than removed here. The divergence is listed as a side effect of the more aggressive continuous-integration optimisations, which create temporary draft pull requests. A supported configuration therefore makes the two records disagree, with no indication in either that the other has been made unreliable.

\subsection{Consequence}\label{sec:ms4.4}

The governing terms allocate consequence uniformly and without reference to the distinctions above: users are made responsible, as against the provider, for evaluating output and for what follows from using it. Section J of GitHub's terms of service, effective April 2026, makes the user responsible for reviewing, testing, and validating any output before use, with indemnity obligations extending to claims arising from output incorporated into products (\eIcode{POL-GH-01}). Cursor's terms place the duty in clause 1.4 (\eIcode{POL-CURSOR-01}), quoted in Section~\ref{sec:ms6.1}, and Sourcegraph's AI Terms of Use make the user solely responsible for reviewing and validating any outputs before use (\eIcode{POL-SG-04}, quoted in Section~\ref{sec:ms6.2}). Anthropic states the duty in both of its instruments, the consumer terms telling the user not to rely on outputs or actions without independently confirming their accuracy (\eIcode{POL-ANTH-01}) and the commercial terms making it the customer's responsibility to evaluate whether outputs are appropriate for the use case (\eIcode{POL-ANTH-02}); JetBrains states it within an enumerated list of user responsibilities (\eIcode{POL-JB-01}). None of them allocates responsibility per event: the duty attaches to output as a class.

Among the documents discussed here, two state no express verification duty: Replit's terms disclaim responsibility for accuracy without assigning a duty to check (\eIcode{POL-REPLIT-01}), and the GitHub terms document governing additional products allocates neither verification nor responsibility for output, delegating both (\eIcode{POL-GH-05}). That is a statement about the documents named here, not a count over the eighteen collected. Figure~\ref{fig:vocabgrid} records that vocabulary across the fourteen of the eighteen that are instruments; the other four allocate nothing, two being index pages (\eIcode{POL-GH-03}, \eIcode{POL-SG-01}), one an amendment log (\eIcode{POL-GH-08}) and one a superseded instrument that states its own deprecation (\eIcode{POL-GH-06}). Two of the eighteen state their own supersession, and only that one falls among the four. The second is Sourcegraph's Cody Enterprise Terms of Use (\eIcode{POL-SG-02}), which allocates, is counted among the fourteen and is the row Figure~\ref{fig:vocabgrid} prints under that title; its opening paragraph confines it to licences granted before two stated dates and adds that ``All other use of Cody is governed by the Sourcegraph Terms''.

\begin{figure}[!htbp]
\centering
\newcommand{\eIpev}[2]{{\scriptsize\textbf{#1}}\newline{\scriptsize #2}}
\setlength{\fboxsep}{0.5mm}
\newcommand{\eIpno}{\fcolorbox[gray]{0.80}{1}{\makebox[11mm][c]{%
  \rule[-0.6ex]{0pt}{2.6ex}\textcolor[gray]{0.45}{\tiny none}}}}
\newcommand{\eIpot}{\fcolorbox[gray]{0.45}{0.82}{\makebox[11mm][c]{%
  \rule[-0.6ex]{0pt}{2.6ex}\tiny other}}}
\setlength{\tabcolsep}{1mm}
\renewcommand{\arraystretch}{1.1}
\begin{tabular}{@{}p{14mm}@{\hspace{1.2mm}}*{9}{>{\raggedright\arraybackslash}p{12.6mm}}@{}}
& \multicolumn{9}{@{}l@{}}{\setlength{\unitlength}{1mm}%
  \begin{picture}(126,3)\put(0,1.2){\vector(1,0){126}}\end{picture}} \\[-1.4mm]
& \eIpev{A.1}{Task assignment} & \eIpev{A.2}{Plan approval}
& \eIpev{A.3}{Change authorship} & \eIpev{A.4}{Pull request creation}
& \eIpev{A.5}{Check execution} & \eIpev{A.6}{Review} & \eIpev{A.7}{Approval}
& \eIpev{A.8}{Merge} & \eIpev{A.9}{Deploy\-ment} \\[1mm]
\hline
\rule{0pt}{3.2ex}\raggedright\scriptsize\textbf{Vendor documentation} &
 {\scriptsize A E R} & {\scriptsize A V R} & {\scriptsize E R} &
 {\scriptsize A E R} & {\scriptsize V} & {\scriptsize V R} &
 {\scriptsize A E V R} & {\scriptsize A V R} & {\scriptsize A V R}
 \tabularnewline[1mm]
\hline
\rule{0pt}{3.4ex}\raggedright\scriptsize\textbf{Provider terms} &
 \eIpot & \eIpno & \eIpot & \eIpno & \eIpno & \eIpno & \eIpot & \eIpno
 & \eIpot \tabularnewline[1mm]
\hline
\multicolumn{10}{@{}p{146mm}@{}}{\rule{0pt}{3.2ex}%
 {\scriptsize\textbf{Consequence} is allocated once per provider, to the user,
 for output as a class, and at no event at all, under rule C-1.}} \\
\hline
\end{tabular}
\caption{The nine events as a pipeline, and what each layer states at each.
\emph{Vendor documentation} gives the dimensions of the map for which the
collected documentation carries an answer at that event, read off
Figure~\ref{fig:eventmap}: A authority, E execution, V verification, R record,
a dimension missing from a cell being one that rule C-2 judges inapplicable
there. \emph{Provider terms} reports the search of all eighteen collected
policy documents for the surface forms naming each event: \emph{none} where
those forms return no occurrence in any of the eighteen, \emph{other} where
they occur in a sense that is not the workflow one. Five of the nine return
none. The vocabulary of the other four
occurs, in no workflow sense: assignment is assignment of the agreement or the
licence, approval is written approval of an assignment, a settlement or a
payment method, commit is a subcontractor's undertaking or a licence-counting
rule, and deployment is service description or software installation. The
terms layer therefore names no workflow event in its workflow sense at any of
the nine.}
\label{fig:pipeline}
\end{figure}

Each of the eighteen collected policy documents was searched for the surface forms naming the nine events of the map, from assign and its inflections through to deploy and deployment; the deposited script is \eIcode{event\_vocabulary\_in\_terms.py} under \eIcode{analysis/\eIbrk{}}. The event vocabulary appears in 21 of the 162 cells and produces 70 occurrences. \textbf{Five of the nine events, plan production and approval, pull request creation, check execution, review, and merge, return no occurrence in any of the eighteen documents}, and reading all seventy occurrences of the other four in context returns none in a workflow sense, as Figure~\ref{fig:pipeline} records. The correspondence RQ1 asks about is therefore not low at the event level but undefined: one of the two layers makes no event-level statement to compare against. The search measures the extent of a premise coding rule C-1 states rather than discovering it, and the two are different claims. C-1 asserts that no collected instrument allocates responsibility at event granularity. What the search establishes is which of the nine events draw no occurrence of their own vocabulary at all, five of them, and in what senses the seventy occurrences the other four draw are used, none of them the workflow one. That second claim is checkable against the deposited script and the archived bodies by a reader who rejects the first, and it is what the result here consists of. Under Section~\ref{sec:ms3.3}'s rule, what is established is that no collected instrument names a workflow event in its workflow sense, not that no duty arises at any event.

\section{Three patterns}\label{sec:Patterns}

Separating Record from the other four dimensions makes an obligation with no corresponding artifact visible; separating Execution from Consequence makes their divergence visible. Neither separation fixes the \emph{direction} or the \emph{extent} of what appears: the map could have shown a record at every event, or execution and consequence coinciding throughout, and shows neither.

\subsection{Responsibility without record}\label{sec:ms5.1}

The terms make verification the user's obligation, and in the documentation collected here no page names an artifact that records its performance, at any of the thirty-six tool-event cells of the map. Every record those pages document records that an action occurred. None records that a check was performed. The page sets behind that scope are named rather than assumed: the audit-log field lists of \eIcode{GH-CTRL-08} for the platform layer and the equalised vendor sets of Section~\ref{sec:ms3.3} for the tools, both read in full and the vendor half read again by the third coder. The claim is of the same form as the eight in that section's absence register and was confirmed by those readings rather than by the register's protocol, so the register's own outcome, three of the eight surviving, is the rate against which to weigh it. One archived page records a reason and scopes itself away from these events in its own words. On GitHub's issue-triage automations ``Every supported action records the reason behind it'', which ``gives you an audit trail of what changed and why'' (\eIcode{EQ-GITHUB-01}); the same page confines the capability to the attributes an automation changes on an issue and states that it does ``not affect other actions an automation can take, such as opening pull requests or pushing code''. What is recorded there is an agent's reason for changing an issue field, not a party's judgement about a change to code, and that boundary is the vendor's own rather than this study's.

\textbf{On the platform's own audit log, an approval and a comment-only review are the same event with the same fields.} The log distinguishes \eIcode{pull\_request\_review.submit} from \eIcode{pull\_request\_review\_comment.create}, but the field list for a submitted review carries no field for the state of that review and none for what the approver assessed (\eIcode{GH-CTRL-08}). A user who reviews an agent's pull request produces an artifact only if they also submit a review. Inspecting and merging without submitting one, which the platform permits unless required approvals are configured, leaves no artifact that the inspection happened. The obligation is asserted in a layer that cannot observe whether it is discharged, and discharged in a layer that records an action and not a check.

This gap has a documented history (Section~\ref{sec:ms1.2}). In every arrangement the code-review studies observed, a person capable of forming the judgement stood at the event and the artifact failed to record what they formed. Figure~\ref{fig:grid}'s agent-occupied cell is the first arrangement in this corpus in which the party at the approval event forms a judgement about the change, is not a person, and emits an approval artifact the record does not distinguish from a person's. Non-persons have withheld merge on a verdict for a decade: required status checks (R14) hold a protected branch until a check passes, and review bots that open the diff have been measured at the review event since Balachandran~\cite{Balachandran2013}. Against Matthias's responsibility-gap condition the documents establish one leg and not the other. The party forming the assessment is not a person and the human's contribution is a threshold set earlier, and the documents say nothing about whether that human could predict what the approver would do, which is what a responsibility gap turns on.

\textbf{The rule-following merge bots.} Automated dependency and merge bots have occupied these events since before agentic tooling, and none of the arrangements they document requires a person to read the diff. GitHub's own instructions for automating Dependabot give a workflow that merges on the update's metadata~\cite{GitHubDependabot2026}; Renovate merges on rules matching package names and update types once the required tests pass~\cite{RenovateAutomerge2026}; Mergify evaluates conditions its users write over a pull request's attributes, labels, checks and approvals~\cite{MergifyConditions2026}; and a widely used auto-approving action approves on a condition such as a match against the acting account~\cite{Marr2026}. These four are published documentation rather than archive items, read on 7 September 2026.

Each applies a rule a person wrote once and did not revisit: the release is a patch, the checks are green. The rule is unconditional with respect to the content of the diff, and no vendor offers it as an assessment of that content. The Cursor product does the other thing. Its documentation states that it ``uses risk scoring, approval policy files, AI review agent findings, and your configuration to decide whether to route or approve'', and that it can ``classify a PR by risk and enforce a maximum approval threshold'' (\eIcode{CURSOR-DOC-08}). The classification is of this change, formed at the event, and the approval is its result. An agent granted approval rights through the same vendor's automations can also approve, request changes and dismiss reviews (\eIcode{CURSOR-DOC-07}), so the judgement it forms can displace one a person formed, and the record marks neither. The boundary therefore runs along the conjunction. Each half alone is old: the auto-approving action emits the artifact on an actor match, and a required status check assesses the change without emitting an approval. What is new is a non-person that forms a content-dependent assessment of the specific change and emits an approval artifact not distinguished from a person's approval in the documented record.

\subsection{Execution separated from consequence}\label{sec:ms5.2}

In the arrangement the terms assume, the party who acts bears the outcome. Across the collected documentation the two separate at three identifiable events. Commit authorship is the first: the agent is recorded as author and the human as co-author while the terms place responsibility on the human, a divergence stated as a traceability feature. This documented separation is consistent with a moral-crumple-zone risk in the sense of Elish~\cite{Elish2019}. What the documentation adds is the mechanism: the boundedness of the nearest human's control is a matter of configuration rather than of attention or skill (Section~\ref{sec:ms5.3}). Rule C-1 codes Consequence as an allocation between provider and user and the user is ordinarily an organisation, so the frame cannot supply the individual bearer that reading needs.

Automatic merging is the second. A pull request configured for auto-merge merges when required reviews and status checks are satisfied, which may be after the approving human has moved on, and is disabled if someone \emph{without} write permissions pushes new changes or switches the base branch, so changes pushed with write access, including by agents on the branch, fall outside the disabling condition (\eIcode{GH-CTRL-09}); the two optional controls that would close it are recorded off at R3 and R4. The third is Cursor's clause 1.7 (Section~\ref{sec:ms6.1}), which assigns the user sole responsibility for the result of a feature that executes code without manual review or confirmation.

\subsection{Optional enforcement}\label{sec:ms5.3}

Thirteen of the fifteen mechanisms that compel are described by their own vendor as something a reader turns on, turns off, or may be permitted to go around. Two are not: Devin's deployment prompt (R18), unconditional wherever the hosted-deployment feature is available though the feature is off for enterprise organisations, and GitHub's constraint at R15, which carries no switch on the archived pages and stands beside an agent-merge affordance the same vendor documents outside the archive (Section~\ref{sec:ms4.1}).

Required approvals apply only where configured (R2); branch protection restrictions do not apply by default to users with admin or bypass permissions (\eIcode{GH-CTRL-01}); rulesets carry explicit bypass lists (\eIcode{GH-CTRL-03}); the gate on workflow execution is removable (R1); on deployment environments the documentation gives the instruction for closing administrator bypass rather than the default (R6); and attribution can be set to empty in three separate settings (Section~\ref{sec:ms4.3}). Two vendors go further. Where a rule permitting only specific commit authors blocks the cloud agent from creating or updating pull requests, GitHub offers as the remedy adding the agent as a bypass actor on the ruleset (\eIcode{GH-AGENT-01}); Cursor's merge queue requires that the repository let the app push or bypass the right merge restrictions (\eIcode{CURSOR-DOC-09}), so the mechanism supplied in place of the control requires rights to push past it. The stronger claim, that a team acquires these tools with the gates already down, is an inference the documents permit but do not make.

\section{Discussion}\label{sec:Discussion}

\subsection{Threats to validity}\label{sec:Threats}

\textbf{Documentation is not practice.} What teams do at these events, and whether the artifacts match observed behaviour, requires repository data.

\textbf{One retrieval date, and documentation that moves under its own citations.} These documents change without notice, and every claim is fixed to the archived text. Between the first retrieval and this revision GitHub renamed Copilot coding agent to Copilot cloud agent and Anthropic moved the Claude Code documentation to a different domain, so paths in the original collection now 404 or redirect, and the archive's record of requested and served URLs is the only reason either change is visible. One change bears on the findings directly. On 1 September 2026 GitHub announced that administrators can authorise Copilot code review to sign off on approval, off by default and configurable at the enterprise, organisation and repository level~\cite{GitHubChangelog2026}. That is a second provider documenting the top-right cell of Figure~\ref{fig:grid}, sixteen days after this study's retrieval date, at the vendor whose archived pages constrain its own agent from approving. Both arrangements are opt-in, and the changelog states its default expressly where Table 3 records R9's as off on this study's inference.

\textbf{Four tools and eighteen policy documents.} The four tools were found by prominence, the scope criterion applied as a test on them rather than as the search that produced them, so no ranking or market measure stands behind the shortlist and OpenAI Codex, Google Jules and Amazon Kiro are in no part of the archive. Twelve of the eighteen policy items, across four providers, govern the four tools studied, Cognition being represented by a website terms of use rather than a product agreement, on which Devin's row in Table 2 rests. The remaining six are the comparison set for Section~\ref{sec:ms6.2}, so no rate is computed over them.

\textbf{Documents outside the collection.} Several documents named in the collected terms were not retrieved, among them the Microsoft product terms the Generative AI Services Terms route to and the acceptable use policies. Each is named in the archive with its referring sentence, and no claim here reaches them. Community conventions were not collected either, so whether practitioners have converged on an agent attribution trailer of their own (Section~\ref{sec:ms4.3}) is a separate question.

\textbf{Two coders reached the same labels on twenty-three of twenty-five decisions, and the two they did not agree on were the same decision seen twice.} The \eIcode{compels} and \eIcode{actor} labels were assigned twice, the second time independently and blind, on a census of the decisions the sheet carried rather than a sample. Cohen's kappa~\cite{Cohen1960} is reported per axis with an exact conditional p obtained by complete enumeration: on \emph{compels}, n = 13, agreement 12 of 13, $\kappa$ = 0.806, 95 per cent confidence interval [0.441, 1.000], p = 0.014; on \emph{actor}, n = 12, agreement 11 of 12, $\kappa$ = 0.750, 95 per cent confidence interval [0.281, 1.000], p = 0.045. Both intervals rest on Cohen's large-sample standard error, whose normal approximation does not hold at this n, so they are to be read as width rather than as calibrated bounds, and no pooled coefficient is reported over two disjoint label sets~\cite{Feinstein1990}. A coefficient computed against the corrected evidence would be 1.00, and it is not reported, because the disagreement is the finding: what the exercise was worth is the audit it forced, not the number it produced.

Both disagreements fell at row R15, where the second coder could not find the quoted constraint in the archive and coded against it. He was right: that page's manifest URL had returned HTTP 404, and the retrieval audit of Section~\ref{sec:ms3.3} followed from his coding. The third coder covered what the second did not, agreeing at R4, R5 and R14 on all three \eIcode{compels} labels and two of three \eIcode{actor} labels, and on the map labels on thirteen of sixteen \eIcode{compels} labels ($\kappa$ = 0.68, exact conditional p = 0.001) and fourteen of sixteen \eIcode{actor} labels ($\kappa$ = 0.77, p < 0.001).

Four conditions bound these coefficients: single coders rather than panels; a second coding obtained after the classification had been drafted, over units the first coder fixed, so that what is tested is the reproducibility of the rules and not the independence of their formation; a third coder's sheets drafted by an AI assistant and ratified by him; and a collection the equal-effort rule raised from 71 items to 121, so that of Appendix A's 116 data cells, the same map counted with the check event's two axis labels in place of a dimension cell, only 44 carry a second reading, while Figure~\ref{fig:vocabgrid}, Table 2's summary, Table 3's \eIcode{Default} column, the 226 extracted passages, the three patterns and the two documented cases carry none.

\textbf{One archived page carries an unusual share of the argument.} \eIcode{CURSOR-DOC-08} supplies the case that forced the two-axis frame (Section~\ref{sec:ms3.2}), the only judgement-forming occupant of Figure~\ref{fig:grid}'s top-right cell, row R9 of Table 3, half of the first case in Section~\ref{sec:ms6.1}, and the thesis of Section~\ref{sec:ms1.2}. Withdraw that page and each loses its exhibit, but not the grid: required status checks at R14 hold the same cell as a compelled check performed by a non-person, so a withdrawal would remove the judgement-forming instance rather than the distinction it was found by.

\textbf{Absences are absences of documentation, and of a particular search.} A tool that documents no merge gate may enforce one; the finding is about what a team reading the vendor's material can determine. Five of this paper's eight absence claims did not survive a reader who had not made them (Section~\ref{sec:ms3.3}), and the three that survived have each been read over the full page set by such a reader.

\textbf{Term counting is not contract interpretation, and the reach of a clause is not settled here.} Section~\ref{sec:ms6.2} counts what parties chose to write down, not what they owe, and Section~\ref{sec:ms6.1} reads how far one clause reaches without deciding it. No governing-law or forum clause was coded for any of the fourteen instruments, background law being outside this study's unit. The sharpest comparison in Section~\ref{sec:ms6.2} runs across two documents rather than a sample. Four of the fourteen instruments of Figure~\ref{fig:vocabgrid}, \eIcode{POL-CURSOR-01}, \eIcode{POL-GH-05}, \eIcode{POL-SG-03} and \eIcode{POL-SG-04}, lie outside the 440-search absence register, their counts coming from the same deposited script over all eighteen items.

\subsection{Implications}\label{sec:Implications}

Each recommendation below is conditional in the way Section~\ref{sec:ms8.1} states: if a team, platform or regulator wants these events answerable after the fact, this is what prevents it. That conditional supplies the forum the corpus does not, so the durability horizons of Section~\ref{sec:ms4.3} become findings about sufficiency only once one of the three is asking.

\textbf{For platforms.} The events that carry obligations should produce artifacts that distinguish them. An approving review is a different act from a comment, and a log recording both as a review submission cannot answer who endorsed a change. Fields already exist to mark an agent actor, and documenting what they mean would let the same log answer whether a change was authored by an agent, by whom it was approved, and whether the two were the same party. That artifact is not the one the runtime-attestation literature is building (Section~\ref{sec:ms2.3}): it would record that the party the terms make responsible formed a judgement before the change merged.

\textbf{For providers}, the vendors in this paper's sense and not the Regulation's (Section~\ref{sec:ms2.1}). Terms written in the vocabulary of input and output describe a relationship the same provider's documentation no longer describes. A definition of output as what a service returns does not reach a service that creates a branch, authors a commit and opens a pull request, and naming the acting party is a precondition for allocating responsibility for the act. A minimal clause would do what clause 1.7 of Cursor's terms half-does for automatic execution (Section~\ref{sec:ms6.1}): identify each mode in which the service performs the approval or executes without confirmation, and state the user's duty in that mode.

\textbf{Two further audiences take the same reading.} For a team or a compliance function, Table 3's \eIcode{Default} column names what would have to be turned on for a stated obligation to be met by configuration rather than by habit: R2 to R5 and R14, every one recorded off on this study's inference; then closing administrator bypass on protected branches and on deployment environments (\eIcode{GH-CTRL-01}, R6) and reading the ruleset's explicit bypass list (\eIcode{GH-CTRL-03}); and, where a vendor's own agent holds approval rights or admits a merge, R9 and R10 record what the repository has to concede. For a regulator, where a coding agent falls within an oversight regime, the nine events are where its measures would have to operate: Article 14(4)'s enabling conditions attach to plan approval, review and approval, and Article 26(2)'s deployer duty to who holds Authority at those events, both conditional on a scope the Regulation does not now extend to these tools (Section~\ref{sec:ms2.1}). Kang~\cite{Kang2026GAIE} prescribes the evidence artifacts an audit would require by tier. One assumption runs under all of this: the instrument is read, if at all, by legal and procurement functions and the documentation by engineers~\cite{Obar2020}.

\subsection{What this paper does not establish}\label{sec:ms8.1}

This study reports a correspondence gap between two published layers. It establishes that the gap exists and where it falls. That it harms anyone is a separate claim this paper does not make. The evidence is documentary, and contains no dispute that turned on a missing artifact, no post-incident review that could not reconstruct who approved a change, no regulatory finding and no team impaired by the arrangement. What the gap does have is a measured footprint. Peer-reviewed repository-scale studies, set out below, count agent-authored pull requests merged with no recorded interaction trace and rejected with no observable decision rationale, which is what the condition mapped here looks like when the artifacts rather than the instruments are the unit of observation.

When the gap would bite follows from the documents rather than from a case. The record of who approved a change is log-durable for one hundred and eighty days (\eIcode{GH-CTRL-07}, \eIcode{EQ-GITHUB-11}), and that log does not distinguish an approval from a comment-only review in any event. A question asked seven months after a merge about who endorsed the change, and on what basis, is therefore not hard to answer from the published record: on these instruments it cannot be answered at all. Incident reviews, audits and disputes are not always opened inside six months.

Two positions hold that the gap is not a defect: that contracts are drafted abstractly because products change faster than instruments can, and that verification is not usefully recordable, a human reading a diff producing no natural trace and a mandated one producing an artifact of compliance rather than of judgement. Neither has to be refuted for the map's finding to hold: an approval artifact does not indicate that a person read the change, and nothing in either layer marks where that is so.

The repository-side measurements this paper cannot make have since been published, and they corroborate the condition the documents describe. Li et al.~\cite{LiAIDev2026} assembled AIDev, a dataset of 932,791 pull requests authored by five agents across 116,211 repositories and 72,189 developers, with a curated subset of 33,596 pull requests from 2,807 repositories above one hundred stars. Peralta et al.~\cite{Peralta2026} took 9,799 human-reviewed agentic pull requests and inspected 717 by hand, reporting among the merged ones 15.4 per cent carrying explicit reviewer involvement and 5.5 per cent showing no visible interaction trace, and among the rejected ones 33.1 per cent carrying no observable decision rationale. Both are peer-reviewed, both were taken at a scale no document collection reaches, and both are consistent with what this paper documents from the instruments: an event that carries an obligation completes, and the artifact it leaves does not record the discharge. The boundary on that support is that these results measure the artifact and not the reading, this paper's own limit reached from the other side: an interaction trace is evidence that someone acted, and its absence is not evidence that no one read the diff. Neither sets its counts against which platform controls were configured at the repositories counted or which provider instrument governed the agent, which is what this paper supplies.

\section{Conclusions}\label{sec:Conclusions}

\textbf{Where the two layers diverge.} Read across nine workflow events, four vendors' documentation and seven providers' terms fail to intersect at the event level. On RQ1 the correspondence is not low but undefined, because five of the nine events return no occurrence of their own vocabulary in any of the eighteen collected instruments, and the seventy occurrences of the other four are not workflow senses. Where both layers speak, the map returns three patterns. Obligations are asserted with no artifact recording their discharge. Execution separates from consequence, the agent being the recorded author and the human the co-author while the instrument allocates responsibility to the human. Enforcement exists, with two documented exceptions, as configuration, so nine of the fifteen compelling mechanisms carry a default this study inferred rather than one a vendor stated. The Record dimension, which answers RQ3, is the thinnest: the artifacts these events produce record that an action occurred, and none of them that a check was performed.

\textbf{What the two-axis decomposition adds.} The study began where the governance frameworks are, reading verification on one scale as enforced, advised or absent, and a case in the corpus could not be placed on it. That scale collapses two independent questions: whether a mechanism withholds progress until the check occurs, and whether the party performing it is a person. Crossing them puts fifteen mechanisms in four cells and makes the top-right cell visible, where at one provider, at the retrieval date, the mechanism compels and the performer is an agent that classifies the change by risk against criteria a human set once, producing an approval the artifact cannot distinguish from a person's judgement on a diff. The same axis places GitHub's required status check, a rule a person wrote once, in that column beside the approver, and what separates the two is what the vendor offers the artifact as (Section~\ref{sec:ms5.1}). The axis records that the performer is not a person, not that a judgement was formed, which is the limit of what the grid establishes.

\textbf{What the two cases show.} Each is internal to a single provider, so neither depends on comparing vendors. In the first, one instrument makes the user responsible for evaluating the use of any suggestion while the same provider's documentation describes a product that performs that evaluation, against criteria the user configured once, and records its result as an approval. In the second, the documentation describes an autonomous agent that authors commits, opens pull requests and is prevented from approving its own work, while the terms describe a return value, in a vocabulary that names no agent; the volume-licence customer, the party most likely to operate agents at scale, carries no express duty to check output in the three documents the provider serves directly, where the individual user carries one. Both are facts about drafting rather than about what either party owes, and Section~\ref{sec:Threats} states that bound.

\textbf{None of these allocations is wrong on its own terms.} Each layer is internally coherent. What the map shows is that the two were built to answer the same question, have not been read against each other, and disagree at the events where agentic software work actually happens. It would be as convenient to conclude that agents broke this as to conclude that agents changed it everywhere, and the evidence supports neither: the approval artifact recorded less than a reader may assume long before agents existed, three of the four tools studied keep a person at the event, and the terms at every provider go on allocating consequence for the output as a class without naming the event or the artifact. What the documentation does establish is narrower and harder to set aside. In every arrangement the code-review studies observed, a party who could have formed a judgement stood at the event. At the retrieval date of 16 August 2026 one vendor shipped a product for which that is not so, and sixteen days later a second documented the same arrangement at the same cell, both opt-in and neither marked in the artifact (Section~\ref{sec:Threats}). The terms allocate consequence against that artifact regardless. The weakness did not appear. It stopped being a matter of how carefully people work and became a matter of what a product is built to do.

\textbf{The successor study is empirical.} It needs repository histories in which the nine events can be located, the approval, merge and log artifacts each event produced, and incident records against which those artifacts can be tested. It would answer, as this paper cannot, what practitioners do at these events and what the correspondence gap costs anyone. The map, the coding frame and the archive are deposited so that the nine events, the five dimensions and the two axes can be carried over unchanged. It also makes a prediction that such a study could refute: the events at which the terms layer is silent and at which no artifact records a check should be the events at which repository histories show the least recorded human involvement, and the events at which a gate is both documented and compelled should show the most. Agent-authored pull request histories of the kind AIDev assembles, joined to each repository's branch-protection and ruleset configuration and to the provider instrument governing the agent, would test that ordering, and no association, or the reverse ordering, would refute it. Nothing in this paper establishes it, which is why it stands here as a prediction and not as a finding.

\section*{Acknowledgements}
The first author thanks the second coder who independently re-coded the classification
reported in Section~\ref{sec:Threats}. That coding produced two disagreements,
both at the same decision, and pursuing them uncovered the retrieval failure
recorded in Section~\ref{sec:ms3.3}.

\section*{Ethics statement}
This study analysed documents published by their providers to the public. It
involved no human subject, collected no personal data, and sought no
ethics-committee approval on that basis. Two people other than the first
author took part. The independent second coder took part voluntarily and
unpaid, asked not to be named and is named in neither this paper, the
Acknowledgements nor any file in the deposit, nothing that identifies him is
recorded in either, and he has confirmed in writing his agreement to the
publication of his notes and to the deposit of his sheet under CC BY 4.0. The
second author coded three sheets from the archive and the written rules before
reading any part of the manuscript, on a written brief deposited with the
coding materials, and joined the paper as co-author after the sheets were
returned. The
findings are dated to the retrieval of 16 August 2026 and are findings about
what providers have documented, not about their conduct.

\section*{CRediT author statement}
Sabry E. Farrag: Conceptualization, Methodology, Software,
Validation, Formal analysis, Investigation, Data curation, Writing -- original
draft, Writing -- review and editing, Project administration. Youssef Saber:
Validation (independent blind coding of three sheets from the archive, and
re-verification of the manuscript's numbers), Formal analysis (the agreement
analysis of that coding), Writing -- review and editing. Software and Writing
-- original draft were performed with the assistance recorded in the
declaration of AI assistance below, which is the fuller statement: the first
author specified the deposited scripts and read, ran and corrected them before
their output entered the study, and revised every drafted passage before it
entered the manuscript. Every coding decision was made, or checked and
adopted, by the author whose name the sheet carries, and each author is
responsible for the parts those roles describe.

\section*{Funding}
This research received no specific grant from any funding agency in the
public, commercial, or not-for-profit sectors.

\section*{Data availability}
The source collection is deposited at Zenodo, concept DOI
\href{https://doi.org/10.5281/zenodo.21965181}{https://doi.org/10.5281/zenodo.21965181},
which resolves to the current version. Section~\ref{sec:ms3.5} gives its
contents and the commands that recompute the reported coefficients; Zenodo publishes
a checksum for the archive file on the record page.

\section*{Competing interests}
The authors declare no competing financial interests or
personal relationships that could have appeared to influence the work reported
in this paper.

\section*{Declaration of generative AI assistance}
During the preparation of this work, the first author
utilised Claude Opus 5 (Anthropic) in order to write and run the retrieval,
extraction, verification and consistency scripts deposited with the archive, to
verify references against publisher sources, and to draft passages of the
text. The second author utilised Claude Sonnet 5 (Anthropic) to sweep the
archived page sets and to draft the labels, verdicts and notes of the three
coding sheets, and Claude Opus 5 to re-check the drafts against the archived
pages, under the brief's rule that every label and verdict returned is
his own decision after checking the archived page. Subsequent to the
application of this tool, the authors reviewed and edited the content as
necessary and accept full responsibility for the final content of the
publication. No language model is present in the retrieval or extraction path:
the scripts issue direct HTTP requests, write response bodies byte-exact and
apply the lexical retrieval rules of the Method, so the archive and the
extraction records are reproducible from the manifest without any model. No
coding decision of the first author was made by a model, and every coding
decision of the second author was checked and adopted by him. The third coding
pass, the blind second coding of the twenty-five decisions, was performed by
neither author: the brief deposited for it required the coder to work from the
evidence quotation and the archived document and to note every hesitation, and
it stated no position on the use of a language model, so this declaration makes
no claim either way about that pass.

\bibliographystyle{IEEEtran_for_EI}

\bibliography{ms}

\appendix
\renewcommand{\thesection}{Appendix~\Alph{section}}
\renewcommand{\thesubsection}{\Alph{section}.\arabic{subsection}}

\section{The map, by tool}\label{sec:A}

The Accountability Event Map is deposited in full as \eIcode{appendix\_A\_evidence\_map.md} (Section~\ref{sec:ms3.5}) and cited throughout this paper by cell, A.1 to A.9. It covers all four tools across all nine events on the five dimensions, in 116 data cells beside the tool column: fifty-six carry a quotation, or a short attribution, with its archive item identifier; thirty-two are the declared \eIcode{compels} and \eIcode{actor} labels of Section~\ref{sec:ms3.2}; twenty are derived labels read off the cited cells in the same row; and eight carry the marker NOT \nohyphens{DOCUMENTED} with the page set searched, seven of them at plan approval and deployment. Figures~\ref{fig:eventmap} to \ref{fig:pipeline} and Tables 1 to 3 are read off those cells. The deposited file also carries the Method material this paper reports in summary: the four units of analysis, the seven steps of the coding procedure, the applicability matrix, the glossary, the retrieval and archive audit, the extraction breakdown, the register of the eight absence claims, the basis for every \eIcode{Default} value in Table 3, the per-document counts behind Figure~\ref{fig:vocabgrid}, the archived pages that corroborate a finding already established from another page, each with its identifier, its served URL and the sentence that carries it, and the reproducibility commands and hashes.

Pages searched per tool after equalisation under the single rule of Section~\ref{sec:ms3.3}, given as archived items and then as distinct documents: GitHub Copilot cloud agent, 24 items over 22 pages; Devin, 14 over 12; Cursor, 32 over 26; Claude Code, 19 over 17. Platform controls applying to all four are \eIcode{GH-CTRL-01} to \eIcode{GH-CTRL-13} and \eIcode{GIT-TRAILER-01}.

\section{The coding frame}\label{sec:B}

The five definitions of Table B1 were fixed before the corpus was first read and did not change; Table B2 is the brief issued to the second coder, the instrument the agreement statistic of Section~\ref{sec:Threats} was produced under. The frame is deposited in full as \eIcode{coding/\eIbrk{}CODEBOOK.md}.

\eIkeepwithcaption{12\baselineskip}
\textbf{Table B1. The five dimensions, with a corpus anchor for each.} A cell is coded where a passage answers the dimension's question; \eIcode{coding/\eIbrk{}CODEBOOK.md} gives the apply and do-not-apply wording for each.

\footnotesize
\def\LTcaptype{eIuncaptioned}
\begin{longtable}{>{\raggedright\arraybackslash}p{6.09cm}>{\raggedright\arraybackslash}p{7.92cm}}
\toprule
\textbf{Dimension} & \textbf{Anchor} \\
\midrule
\endhead
\textbf{Authority.} \emph{Who is permitted to cause this event.} & ``Privilege escalation controls: The cloud agent only responds to interactions from users with repository write access.'' (\eIcode{GH-AGENT-06}) \\
\textbf{Execution.} \emph{Who performed it, and under which recorded identity.} & ``Commits from Copilot cloud agent are authored by Copilot, with the person who started the task listed as co-author.'' (\eIcode{GH-AGENT-05}) \\
\textbf{Verification.} \emph{Who is obliged to check, whether a mechanism compels the check, and which party performs it.} & Compelled: ``Actions workflows triggered by pull requests raised by the agent require approval from a user with write access before they will run.'' (\eIcode{GH-AGENT-06}). Not compelled: ``review Claude's changes before merging.'' (\eIcode{CC-DOC-02}) \\
\textbf{Consequence.} \emph{Which party the governing instrument allocates responsibility for the output to, as between provider and user.} Coded once per provider under rule C-1, from the policy instruments only. & ``You are solely responsible for reviewing and validating any Outputs before use.'' (\eIcode{POL-SG-04}) \\
\textbf{Record.} \emph{Which artifact shows that the event occurred,} and not a transient interface state. & the field \eIcode{actor\_is\_agent}, documented as ``Indicates whether the actor is an AI agent.'' (\eIcode{EQ-GITHUB-08}) \\
\bottomrule
\end{longtable}
\def\LTcaptype{table}
\normalsize

\eIkeepwithcaption{12\baselineskip}
\textbf{Table B2. The two axes of Verification, as issued to the second coder.}

\footnotesize
\def\LTcaptype{eIuncaptioned}
\begin{longtable}{>{\raggedright\arraybackslash}p{1.69cm}>{\raggedright\arraybackslash}p{6.52cm}>{\raggedright\arraybackslash}p{5.39cm}}
\toprule
\textbf{} & \textbf{\eIcode{compels}} & \textbf{\eIcode{actor}} \\
\midrule
\endhead
Permitted labels & \eIcode{Yes}, \eIcode{No} & \eIcode{Human}, \eIcode{Agent} \\
Decision rule, as issued & ``The test is not whether the thing is important, or whether a careful team would do it. The test is whether the product refuses to proceed without it.'' & ``Read the quoted text closely here. Some passages describe an agent performing an action that is conventionally a human's.'' \\
Boundary, fixed in advance & A mechanism that compels but can be turned off, or bypassed by an administrator, is still \eIcode{Yes}; whether it is on by default is a separate matter. Rule B-1 extends this to a gate inside an optional operating mode. & The binary is non-exhaustive: it has no cell for a party that is neither a person nor an AI coding agent, and Cursor's merge queue is forced to \eIcode{Agent} by the label set. \\
\bottomrule
\multicolumn{3}{@{}p{\dimexpr\textwidth-2\tabcolsep\relax}@{}}{\footnotesize \textbf{Note.} The archived document is the source of truth rather than what a coder knows of how a product behaves, and abstention is forbidden. In the map, \eIcode{compels} admits a third value, \eIcode{Not documented}: \eIcode{No} where the pages document the event and nothing is withheld at it, \eIcode{Not documented} where they carry no statement of the kind at all.}
\end{longtable}
\def\LTcaptype{table}
\normalsize

\end{document}